\documentclass[fleqn,usenatbib]{mnras}

\usepackage{newtxtext,newtxmath}

\usepackage[T1]{fontenc}

\DeclareRobustCommand{\VAN}[3]{#2}
\let\VANthebibliography\thebibliography
\def\thebibliography{\DeclareRobustCommand{\VAN}[3]{##3}\VANthebibliography}

\usepackage{CJK}
\usepackage{graphicx}	
\usepackage{amsmath}	
\usepackage{bm}

\newcommand{\bHI}{b_{\mathrm{H\textsc{i}}}}
\newcommand{\bg}{b_{\mathrm{g}}}
\newcommand{\Tb}{\overline{T}_{\mathrm{H\textsc{i}}}}
\newcommand{\hi}{H\textsc{i}}
\newcommand{\secref}[1]{\hyperref[#1]{Section~\ref*{#1}}}
\newcommand{\appref}[1]{\hyperref[#1]{Appendix~\ref*{#1}}}

\defcitealias{2026arXiv260701864C}{Chen and Cunnington et~al.}
\newcommand{\citetaliasyear}[1]{\citetalias{#1}~(\citeyear{#1})}  
\newcommand{\citepaliasyear}[1]{(\citetalias{#1}~\citeyear{#1})}   

\makeatletter
\patchcmd{\ps@titlepage}
  {\def\@oddfoot{\small\copyright\ \@pubyear\ The Authors\hfil}}
  {\def\@oddfoot{%
     \small\copyright\ \@pubyear\ The Authors\hfill
     \begin{minipage}[b]{\columnwidth}%
       \raggedright\footnotesize
       \hypertarget{corautga}{\textcolor{\@linkcolor}{$^{\star}$}}\hspace{0.35em}E-mail: gautieri@sissa.it\\
       \hypertarget{corautms}{\textcolor{\@linkcolor}{$^{\dagger}$}}\hspace{0.35em}E-mail: marta.spinelli@oca.eu\\
       \hypertarget{corautzc}{\textcolor{\@linkcolor}{$^{\ddagger}$}}\hspace{0.35em}E-mail: zhaoting.chen@roe.ac.uk
     \end{minipage}}}
  {}{\errmessage{Could not place corresponding-author emails}}
\makeatother
\newcommand{\correspondingemails}{\clearpage}

\title[Towards constraining cosmology with MeerKLASS]{Towards constraining cosmology with HI--galaxy cross-correlations from MeerKLASS}

\author[G. Autieri et al.]{Gabriele Autieri,$^{1, 2, 3}$\textsuperscript{\hyperlink{corautga}{$\star$}}
Marta Spinelli,$^{4,5}$\textsuperscript{\hyperlink{corautms}{$\dagger$}}
Zhaoting Chen (陈兆庭),$^{6,4}$\textsuperscript{\hyperlink{corautzc}{$\ddagger$}}
Matteo Viel,$^{1,2,3,7,8}$
\newauthor
Matilde Barberi-Squarotti,$^{9, 10, 11}$
Phil Bull,$^{12}$
Stefano Camera,$^{13, 14, 15, 16}$
Isabella P. Carucci,$^{3,7}$
\newauthor
Steven Cunnington,$^{17}$
Jos\'e Fonseca,$^{18, 19, 5}$
Keith Grainge,$^{12}$
Melis O. Irfan,$^{20}$
Yichao Li,$^{21}$
Alkistis Pourtsidou,$^{6,22}$
\newauthor
Mario G. Santos,$^{5}$
Jingying Wang (王婧颖),$^{23, 5}$
and Laura Wolz$^{12}$
\\
\\
$^{1}$SISSA-- International School for Advanced Studies, Via Bonomea 265, 34136 Trieste, Italy\\
$^{2}$INFN-- National Institute for Nuclear Physics, Via Valerio 2, I-34127 Trieste, Italy\\
$^{3}$IFPU, Institute for Fundamental Physics of the Universe, via Beirut 2, 34151 Trieste, Italy\\
$^{4}$Observatoire de la C\^ote d'Azur, Laboratoire Lagrange, Bd de l'Observatoire, CS 34229, 06304 Nice cedex 4, France\\
$^{5}$Department of Physics and Astronomy, University of the Western Cape, Robert Sobukwe Road, Cape Town 7535, South Africa\\
$^{6}$Institute for Astronomy, The University of Edinburgh, Royal Observatory, Edinburgh EH9 3HJ, UK\\
$^{7}$INAF, Osservatorio Astronomico di Trieste, Via G. B. Tiepolo 11, I-34131 Trieste, Italy \\
$^{8}$ICSC - Italian Research Center on High Performance Computing, Big Data and Quantum Computing, via Magnanelli 2, 40033, Casalecchio di Reno, Italy \\
$^{9}$Universit\`a degli Studi di Milano, Via Celoria 16, 20133 Milan, Italy\\
$^{10}$INAF -- Istituto Nazionale di Astrofisica, Osservatorio Astrofisico di Brera-Merate, Via Brera 28, 20121, Milan, Italy\\
$^{11}$INFN -- Istituto Nazionale di Fisica Nucleare, Sezione di Milano, Via Celoria 16, 20133 Milan, Italy\\
$^{12}$Jodrell Bank Centre for Astrophysics, Department of Physics \& Astronomy, The University of Manchester, Manchester M13 9PL, UK \\
$^{13}$Dipartimento di Fisica, Universit\`a degli Studi di Torino, Via P.\ Giuria 1, 10125 Torino, Italy\\
$^{14}$INFN -- Istituto Nazionale di Fisica Nucleare, Sezione di Torino, Via P.\ Giuria 1, 10125 Torino, Italy\\
$^{15}$INAF -- Istituto Nazionale di Astrofisica, Osservatorio Astrofisico di Torino, Strada Osservatorio 20, 10025 Pino Torinese, Italy\\
$^{16}$Department of Physics \& Astronomy, University of the Western Cape, 7535 Cape Town, South Africa\\
$^{17}$Institute of Cosmology \& Gravitation, University of Portsmouth, Dennis Sciama Building, Portsmouth, PO1 3FX, UK \\
$^{18}$Instituto de Astrofisica e Ci\^{e}ncias do Espa\c{c}o, Universidade do Porto CAUP, Rua das Estrelas, PT4150-762 Porto, Portugal\\
$^{19}$Departamento de F\'isica e Astronomia, Faculdade de Ci\^{e}ncias, Universidade do Porto, Rua do Campo Alegre 687, 4169-007 Porto, Portugal\\
$^{20}$Institute of Astronomy, University of Cambridge, Cambridge, CB3 0HA\\
$^{21}$Department of Physics, College of Sciences, Northeastern University, Wenhua Road, Shenyang 11089, China\\
$^{22}$Higgs Centre for Theoretical Physics, School of Physics and Astronomy, Edinburgh EH9 3FD, UK\\
$^{23}$State Key Laboratory of Radio Astronomy and Technology, Shanghai Astronomical Observatory, CAS, 80 Nandan Road, Shanghai 200030, China
}

\date{Accepted XXX. Received YYY; in original form ZZZ}

\pubyear{\the\year{}}

\begin{document}
\begin{CJK*}{UTF8}{gbsn}
\label{firstpage}
\pagerange{\pageref{firstpage}--\pageref{lastpage}}
\maketitle

\begin{abstract}

We present an inference of cosmological parameters from the cross-correlation power spectrum between the MeerKLASS 2019 L-band \hi\ intensity maps and the overlapping WiggleZ galaxies at $z\simeq0.43$. 
The analysis is intended as a proof of concept to verify that the full pipeline, from the foreground cleaning of the maps to the likelihood analysis, can be applied end-to-end to real data and can return results consistent with the concordance cosmology. 
The model of the observed power spectrum includes observational effects that are relevant to a single-dish experiment. 
In the most optimistic scenario in which we hold the astrophysical nuisance parameters fixed, we obtain $H_0=65.3^{+5.8}_{-10}\;\mathrm{km\,s^{-1}\,Mpc^{-1}}$ and $\Omega_\mathrm{m}=0.44^{+0.12}_{-0.30}$ at the $68\%$ confidence level. 
We find that sampling the \hi\ amplitude with an informative prior retains most of the constraining power and changes these results by less than $1\sigma$. 
Moreover, using synthetic data, we show that neglecting the observational effects in the model biases the inferred parameters by up to $2.5\sigma$ and also underestimates their uncertainties. 
Finally, we show that varying the number of foreground modes removed and the range of scales included changes the inferred parameters by less than $0.5\sigma$. 
These results demonstrate the robustness of our pipeline in modelling \hi\ intensity mapping data, and set the foundation for future cosmological inference with intensity mapping using MeerKLASS and SKAO.
\end{abstract}

\correspondingemails



\section{Introduction}
\label{sec:intro}

The 21\,cm line produced by the hyperfine transition of neutral hydrogen (\hi) offers a direct way of tracing the large-scale structure of the Universe over large volumes and a wide range of redshifts. Rather than resolving individual galaxies, intensity mapping (IM) measures the integrated 21\,cm emission of many unresolved sources within each resolution element, treating the brightness temperature field as a biased tracer of the underlying matter distribution \citep{2013MNRAS.434.1239B, Bull:2014rha}. Since the signal is spectroscopic by construction, the radial coordinate is obtained at essentially no additional cost, and large cosmological volumes can be surveyed with comparatively modest integration times. This makes the technique an attractive complement to optical galaxy surveys, and it is one of the cosmological science drivers of the Square Kilometre Array Observatory (SKAO) \citep{2020PASA...37....7S, Wolz:2026ska}.

The main challenge for 21\,cm intensity mapping is the foreground emission. Galactic synchrotron and free-free emission, together with extragalactic point sources, exceed the cosmological signal by four to five orders of magnitude. Their spectral smoothness makes it possible, in principle, to separate them from the 21\,cm fluctuations, and blind component separation methods based on the frequency--frequency covariance of the maps have been shown to be effective at doing so \citep{Wolz:2013wna, Alonso:2014dhk}. The separation is, however, never clean: the same smoothness that identifies the foregrounds is shared by the \hi\ modes with low radial wavenumber, which are removed together with them. Moreover, any effect that perturbs the spectral structure of the data, such as calibration errors, polarisation leakage and beam chromaticity, leaks contamination into the modes that remain \citep{Soares:2020zaq, Cunnington:2020mnn, Matshawule:2020fjz}. The resulting signal loss is substantial and must be corrected for, and the residual contamination is additive and not straightforward to characterise \citep{Cunnington:2023ftf, Carucci:2024qpm}.

Cross-correlation with an overlapping galaxy catalogue mitigates both problems. Foreground residuals, thermal noise and instrumental systematics in the intensity maps are uncorrelated with the positions of optically selected galaxies, so they contribute to the variance of the cross-power spectrum but not to its expectation value \citep{Wolz:2015ckn}. In the auto-power spectrum, by contrast, these terms enter as an additive bias that has to be modelled or subtracted, which is considerably more demanding. This is why the first and most robust measurements of 21\,cm clustering have all been obtained in cross-correlation \citep{Chang:2010jp, Masui:2012jk, Switzer:2013ewa, Anderson:2017ert, eBOSS:2021ebm, CHIME:2022kvg}, while auto-power measurements have followed only later and remain harder to interpret \citep{Switzer:2013ewa, Paul:2023yrr, 2025arXiv251119620C, MeerKLASS:2026ztn}. Using two tracers of the same volume also has advantages that go beyond the suppression of additive systematics: the ratio of tracers with different bias is not affected by cosmic variance in the same way as either tracer alone, which allows some parameter combinations to be constrained beyond the limit set by the number of modes in the survey \citep{Seljak:2008xr, McDonald:2008sh}. Multi-tracer techniques of this kind are by now part of the standard analysis of galaxy surveys \citep{2024MNRAS.532..783Z, DESI:2025yuk}, and intensity mapping is a natural setting for them, since the 21\,cm field and an optical catalogue trace the same structures with very different biases and, more importantly, with entirely different systematics.

Among current experiments, MeerKAT, a 64-dish array in the Karoo desert and a precursor to the SKAO, has produced the most advanced results in single-dish intensity mapping. Operating the dishes in autocorrelation mode rather than as an interferometer sacrifices angular resolution but allows fast scanning of large sky areas and measurements of the large angular scales,
therefore recovering the linear cosmological scales that carry most of the information \citep{2013MNRAS.434.1239B}. The MeerKAT Large Area Synoptic Survey (MeerKLASS) \citep{MeerKLASS:2017vgf} has used this strategy in a pilot survey in L-band \citep{MeerKLASS:2021ivf}, in a deeper L-band field \citep{2025MNRAS.537.3632M}, and, currently, in an ongoing UHF-band campaign aiming at $\sim10{,}000\;\mathrm{deg^2}$ over $0.4<z<1.45$ \citep{Cunnington:2026opd}. The pilot data have yielded a detection of the cross-correlation power spectrum with the WiggleZ Dark Energy Survey at $7.7\sigma$ \citep{Cunnington:2022uzo}, subsequently confirmed with an independent treatment of the foreground separation \citep{Carucci:2024qpm}. The deeper L-band field has since provided a further measurement of the cross-correlation power spectrum \citep{2025MNRAS.537.3632M} with the Galaxy And Mass Assembly (GAMA) survey \citep{2009A&G....50e..12D, 2022MNRAS.513..439D}, a detection of the \hi\ signal through emission-line stacking on galaxy positions \citep{2025ApJS..279...19C} and, more recently, a measurement of the \hi\ auto-power spectrum \citep{MeerKLASS:2026ztn}.

These measurements are usually quoted as constraints on the amplitude combination $\Omega_{\mathrm{\hi}}\bHI r$, which is what the amplitude of the cross-power spectrum probes once a cosmology is assumed. The step of turning the same measurements into constraints on cosmological parameters has not yet been taken, although the modelling required to do so has been developed and studied in some detail in the forecasting literature, both for the \hi\ auto-power spectrum and for its cross-correlation with galaxy surveys \citep{Obuljen:2022cjo, Berti:2022ilk, Casas:2022vik, Berti:2023viz, Autieri:2025sxz}. Extracting cosmological information from these data requires an accurate description of the anisotropic observational effects that reshape the measured power spectrum. In particular, the finite angular resolution of the telescope beam suppresses small-scale transverse modes, i.e. modes with large $k_\perp$, whereas foreground cleaning predominantly removes the spectrally smooth, long-wavelength modes along the line of sight, corresponding to small $k_\parallel$. Pixelisation, gridding, and other geometrical effects introduced by the analysis further modify the observed clustering signal. Accurate modelling of these effects is therefore essential, as residual scale- and angle-dependent distortions can bias cosmological parameter inference. Quantifying these biases, and verifying that the analysis can be run end-to-end without being limited by them, are the purposes of this work. We use the cross-correlation power spectrum monopole measured from the MeerKLASS 2019 L-band pilot data and the overlapping WiggleZ galaxies to infer cosmological parameters. We perform the data analysis and the theoretical modelling with the \textsc{meer21cm} pipeline \citepaliasyear{2026arXiv260701864C}, implemented as a custom likelihood  in the sampler \textsc{Cobaya} \citep{Torrado:2020dgo}. We regard the result as a proof of concept, and we present it as such: the data set covers a narrow redshift range and has limited depth, and the constraints we obtain are correspondingly only sufficient for validation purposes. What we aim to show is that the inferred parameters are consistent with the concordance cosmology and that the pipeline behaves as expected when the underlying assumptions are varied. Accordingly, a large part of this paper is devoted to quantifying how much each assumption contributes and to verifying that the result is stable against the choices made in the data analysis.

The paper is organised as follows. \secref{sec:data} describes the intensity maps and the galaxy catalogue. \secref{sec:methodology} presents the data analysis, from the foreground cleaning to the estimation of the cross-power spectrum and the transfer function correction, and the modelling of the observed power spectrum, including the observational effects specific to single-dish intensity mapping. \secref{sec:results} presents our reference cosmological constraints together with the tests of the modelling assumptions, and \secref{sec:robustness} the tests of the robustness of the result against the data analysis choices. We conclude in \secref{sec:conclusions}.

\section{Data}
\label{sec:data}

The data used in this work have already been presented and analysed elsewhere, and we limit ourselves here to the information relevant for our analysis, referring to \citet{MeerKLASS:2021ivf}, \citet{Cunnington:2022uzo} and \citet{Carucci:2024qpm} for a complete description.

The intensity maps come from the MeerKLASS pilot survey carried out with MeerKAT in 2019 \citep{MeerKLASS:2021ivf, Cunnington:2022uzo}. MeerKAT consists of $64$ dishes of $13.5\,$ m diameter located in the Karoo desert in South Africa. For MeerKLASS intensity mapping, the array is not used as an interferometer. Instead, each dish operates as an independent single-dish telescope, recording the autocorrelation of its voltage signal. This preserves sensitivity to the large angular scales that carry most of the cosmological signal, at the cost of an angular resolution set by the size of a single dish, $\theta_{\mathrm{FWHM}}\simeq1.48^\circ$ at the frequencies used here. 

The observations were taken over six nights between February and July 2019, for a total of $10.5\,$ hours of integration, and targeted the 11hr field of the WiggleZ Dark Energy Survey, covering approximately $200\;\mathrm{deg^2}$. We use the L-band data in the frequency range $971<\nu<1023\,$ MHz, corresponding to $0.388<z<0.463$, with an effective redshift $z_{\mathrm{eff}}\simeq0.425$ for our measurement. The frequency channels have a width of $\Delta\nu\simeq 209\,$kHz. The calibration and the map-making follow the pipeline described in \citet{MeerKLASS:2021ivf}, which includes the flagging of radio frequency interference, the removal of the channels most affected by it, and the projection of the calibrated time-ordered data onto sky maps in the $(\mathrm{R.A.},\mathrm{Dec.},\nu)$ coordinate system, with a pixel size of $0.3^\circ$. Together with the maps, the pipeline provides the number of time samples that have contributed to each map voxel, which we use as a weight for the gridding of the sky map voxels into a Cartesian grid. The maps are dominated by the continuum foreground emission, at the level of a few kelvin, as shown in the top panel of \autoref{fig:maps}. The gridding process and the removal of foregrounds are described later in \secref{subsec:data_analysis}.

For the cross-correlation we use the galaxies of the WiggleZ Dark Energy Survey \citep{Drinkwater:2009sd, Drinkwater:2018wbk}, a spectroscopic survey of UV-selected emission-line galaxies carried out with the Anglo-Australian Telescope. We use the 11hr region, which is the one overlapping with the intensity maps, and retain the $4031$ galaxies that fall within the sky area and the redshift range of the maps, as in \citet{Cunnington:2022uzo}. The redshift uncertainty of the catalogue is negligible for the scales considered here. Following the analysis of the WiggleZ auto-power spectrum by \citet{2011MNRAS.415.2876B}, we adopt a linear galaxy bias $\bg=0.911$ at the effective redshift of the measurement, and we treat it as known in the inference, for the reasons discussed in \secref{sec:results}.

Since the cross-correlation is only sensitive to the volume common to the two data sets, both are restricted to the same region before the analysis, as described in \secref{subsec:data_analysis} and illustrated in \autoref{fig:maps}. Previous analyses of this same data set have reported a detection of the cross-correlation power spectrum at $7.7\sigma$, corresponding to $\Omega_{\mathrm{\hi}}\bHI r=[0.86\pm0.10\,(\mathrm{stat})\pm0.12\,(\mathrm{sys})]\times10^{-3}$ at $k_{\mathrm{eff}}\simeq0.13\;h\,\mathrm{Mpc^{-1}}$ \citep{Cunnington:2022uzo}, and, with a different approach to the foreground separation, $\Omega_{\mathrm{\hi}}\bHI r=[0.93\pm0.17]\times10^{-3}$ \citep{Carucci:2024qpm}. These values provide a useful reference against which to check the amplitude of the signal recovered by our own analysis, as we do in \secref{subsec:planck_cross}.

\section{Methodology}
\label{sec:methodology}

\subsection{Data analysis}
\label{subsec:data_analysis}

\begin{figure}
\centering
\includegraphics[width=\columnwidth]{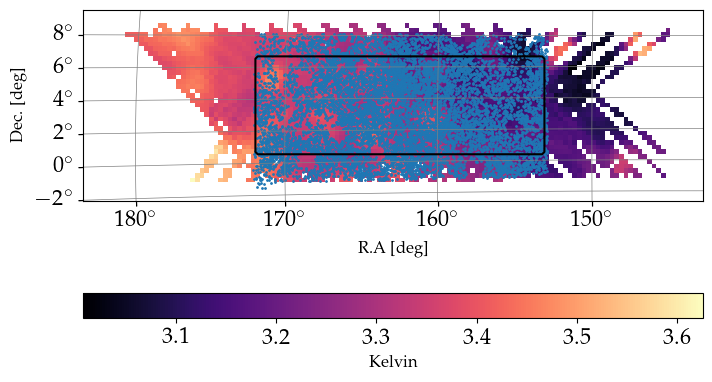}
\includegraphics[width=\columnwidth]{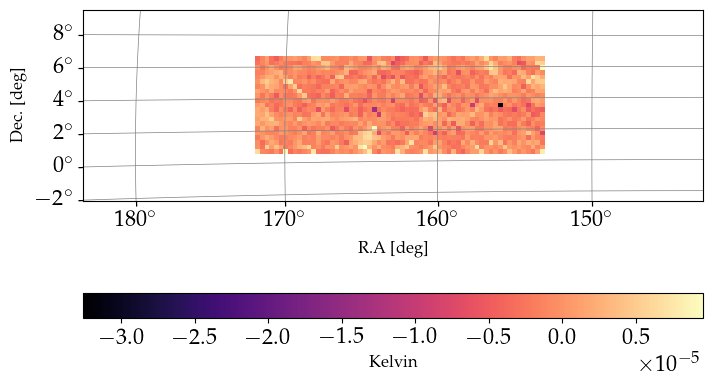}
\caption{\label{fig:maps}
\textit{Top:} \hi\ intensity map before foreground cleaning, with the positions of the overlapping galaxies shown as points. The black rectangle marks the region, $153^\circ<\mathrm{R.A.}<172^\circ$ and $0.5^\circ<\mathrm{Dec.}<6.6^\circ$, to which both data sets are trimmed before the cross-correlation is computed. The map is dominated by the foreground emission, at the level of a few kelvin. \textit{Bottom:} the same map after the removal of $N_{\mathrm{fg}}=10$ PCA modes and the trimming to the common region. The amplitude of the residual field is smaller by about five orders of magnitude.}
\end{figure}

The data analysis is performed with \textsc{meer21cm}\footnote{\url{https://meer21cm.readthedocs.io}} \citepaliasyear{2026arXiv260701864C}, a publicly available \textsc{python} package developed within the MeerKLASS collaboration for the analysis of single-dish \hi\ intensity mapping surveys. The package is survey-oriented: the survey footprint, frequency range and data analysis settings are specified once and are then propagated consistently through every step of the analysis, so that the same choices enter the power spectrum estimation, the transfer function calculation and the theoretical model. This consistency is important here, since, as discussed in \secref{subsec:model}, several of the corrections entering the model of the observed power spectrum depend explicitly on the geometry of the maps and on the gridding settings. We summarise below the steps of the analysis and the choices we have made; the formalism follows \citet{Cunnington:2022uzo} and \citetaliasyear{2026arXiv260701864C}, to which we refer for a more detailed description.

\subsubsection*{Input data and common volume}

The analysis takes as input the \hi\ intensity maps, in the form of a temperature map $m(\boldsymbol{\theta},\nu)$ defined on a grid of sky coordinates $\boldsymbol{\theta}=(\mathrm{R.A.},\mathrm{Dec.})$ and frequency $\nu$, together with a map of the number of hits per voxel, which we use as the map-level weight $w_{\mathrm{\hi}}(\boldsymbol{\theta},\nu)$, and the positions $(\mathrm{R.A.},\mathrm{Dec.},z)$ of the overlapping galaxies. The two data sets cover different and partially overlapping regions of the sky; since the cross-correlation only receives a contribution from the common volume, we trim both to the rectangular region $153^\circ<\mathrm{R.A.}<172^\circ$, $0.5^\circ<\mathrm{Dec.}<6.6^\circ$, shown by the black rectangle in the top panel of \autoref{fig:maps}. The telescope beam is modelled as a Gaussian whose width is computed from the full width at half maximum $\theta_{\rm{FWHM}}\simeq 1.48^\circ$, as described in \secref{subsec:model}.

\subsubsection*{Foreground cleaning}

The maps are dominated by continuum foreground emission, which exceeds the \hi\ signal by several orders of magnitude. However, it is, to a good approximation, spectrally smooth and therefore confined to a small number of modes of the frequency--frequency covariance. We remove it with a blind principal component analysis (PCA) cleaning. After mean-centring the map in each frequency channel, the
frequency--frequency covariance is computed as
\begin{equation}
    \mathsf{C}_{\nu_1\nu_2} = \frac{\sum_{i}\bigl[(w m)(\boldsymbol{\theta}_i,\nu_1)\times (w m)(\boldsymbol{\theta}_i,\nu_2)\bigr]}{\sum_{i} w(\boldsymbol{\theta}_i,\nu_1)\,w(\boldsymbol{\theta}_i,\nu_2)},
    \label{eq:freq_covariance}
\end{equation}
where $w$ is the weighting for the intensity map voxels and the sum runs over the pixels of the angular plane. 
In practice, the weighting for the covariance calculation can be uniform or correspond to inverse noise variance.
We use uniform weighting to be consistent with \cite{Cunnington:2022uzo} and refer readers to \cite{Carucci:2024qpm} for a discussion.
The eigenvectors of $\mathsf{C}$, $\boldsymbol{v}_n$, ordered by decreasing eigenvalue, define the projection matrix
\begin{equation}
    \mathsf{R}^{\mathrm{PCA}} = \mathsf{I} - \sum_{n=1}^{N_{\mathrm{fg}}} \boldsymbol{v}_n \boldsymbol{v}_n^{\mathrm{T}},
    \label{eq:pca_matrix}
\end{equation}
and the cleaned map is obtained as $m_{\mathrm{clean}}(\boldsymbol{\theta},\nu_i)=\sum_j \mathsf{R}^{\mathrm{PCA}}_{ij}\, m(\boldsymbol{\theta},\nu_j)$. In the reference analysis we remove $N_{\mathrm{fg}}=10$ modes; the robustness of the inference to varying $N_{\mathrm{fg}}$ is examined in \secref{subsec:nfg}. The effect of the cleaning is illustrated in \autoref{fig:maps}: the residual field has an amplitude of order $10^{-5}$ K, to be compared with the few kelvin of the uncleaned map, and no large-scale structure associated with the foregrounds is visible by eye. The cleaning inevitably removes part of the \hi\ signal as well, in particular the modes with low $k_\parallel$, which are the ones most easily confused with a smooth spectral component. This signal loss is corrected for at the level of the power spectrum, as described below.

\subsubsection*{Regridding onto a Cartesian grid}

The power spectrum is estimated with fast Fourier transforms and therefore requires both tracers to be defined on a regular Cartesian grid, whereas the intensity maps are defined in sky coordinates and the galaxies are a discrete set of angular positions and redshifts. We assume a fiducial cosmology to convert $(\boldsymbol{\theta},\nu)$ into comoving coordinates $\boldsymbol{x}$, and determine the minimum rectangular box in comoving space that encloses the survey lightcone, adding a small buffer in the transverse directions to avoid the footprint touching the box boundaries. The resulting box has sides of $230.9\times629.0\times283.3\;\mathrm{Mpc}$, sampled with $17\times47\times17$ cells; the cell size is chosen to be coarser than the native resolution of the maps, so that the discreteness of the sampling procedure described below does not dominate the smallest scales retained in the analysis. 

For the galaxies, the assignment is straightforward, since the catalogue is a discrete sample and each object can be placed directly in the cell containing it. For the intensity maps the situation is different: each map voxel is an extended region of $(\boldsymbol{\theta},\nu)$ space whose shape, once transformed into comoving coordinates, is neither Cartesian nor of constant size, so that a direct assignment of the voxel centres to cells introduces a spurious, position-dependent smoothing and aliases power. We therefore follow the resampling scheme of \citet{Cunnington:2023aou}, in which each map voxel is populated with $N_{\mathrm{part}}=20$ sampling particles, distributed uniformly within the voxel and each carrying the temperature of the voxel divided by the number of particles. The particles are transformed to Cartesian coordinates and assigned to the grid with a nearest-grid-point scheme. The number of particles falling in each cell defines the grid-level weight $w^g_{\mathrm{\hi}}(\boldsymbol{x})$ of the \hi\ field, which also encodes the survey selection: cells outside the lightcone receive no particles and are excluded. For the galaxy field we use the binary mask of the same footprint as the field-level weight, so that the two tracers are defined over the same volume, and we convert the galaxy counts into an overdensity by mean-centring and normalising by the mean count within the footprint. The \hi\ field is already mean-centred by the PCA cleaning and is left in temperature units.

\subsubsection*{Power spectrum estimation}

The cross-power spectrum is estimated as
\begin{equation}
    \hat{P}_\mathrm{H\textsc{i},g}(\boldsymbol{k}) = \frac{\mathrm{V}\,\bigl\langle w^g_{\mathrm{g}}\bigr\rangle_{\mathrm{V}}}{\bigl\langle w^g_{\mathrm{g}}w^g_{\mathrm{\hi}}\bigr\rangle_{\mathrm{V}}}\,\Bigl|\bigl[\tilde{\delta}^{\mathrm{n}}_{\mathrm{obs}}(\boldsymbol{k})\bigr]\bigl[\tilde{T}_{\mathrm{obs}}(\boldsymbol{k})\bigr]^{*}\Bigr|_{\mathrm{Re}},
    \label{eq:cross_estimator}
\end{equation}
where $\mathrm{V}$ is the volume of the box, the tildes denote Fourier transforms of the weighted fields, $\langle\rangle_{\mathrm{V}}$ is a volume average, the superscript $g$ denotes the grid-level weights, and only the real part of the product is retained, the imaginary part being consistent with noise. Before the transform, a taper is applied to both fields along the line of sight, in order to suppress the leakage of power caused by the finite radial extent of the box. The three-dimensional power spectrum is then averaged into one-dimensional bins. Before averaging we exclude the regions of the $(k_\perp,k_\parallel)$ plane that are least reliable, retaining only the modes with $0.02<k_\perp,k_\parallel<0.25\;h\,\mathrm{Mpc^{-1}}$. As previously discussed, the lowest wavenumbers are the most affected by the foreground cleaning in the radial direction and by the finite footprint in the transverse one, while the highest $k_\perp$ modes are strongly suppressed by the beam. The impact of this choice is quantified in \secref{subsec:scale_cuts}. The surviving modes are averaged into $11$ linearly spaced bins between $k=0.07$ and $0.3\;h\,\mathrm{Mpc^{-1}}$. For each bin we record the effective wavenumber, i.e. the weighted mean of $|\boldsymbol{k}|$ over the modes entering the bin, which is the wavenumber at which the model is evaluated.

\subsubsection*{Transfer function correction}

The final step of the analysis is the correction for the signal loss caused by the foreground cleaning. We follow the mock signal injection procedure of \citet{Cunnington:2022uzo}, as implemented in \textsc{meer21cm}. A mock \hi\ field $m_{\mathrm{mock}}$, generated with the same survey specifications and a fiducial clustering amplitude, is added to the uncleaned map. Then the PCA is re-run on $m+m_{\mathrm{mock}}$, and the resulting projection matrix is applied to the mock alone, so that the residual $m_{\mathrm{clean,mock}}$ isolates the effect of the cleaning on a known input. The transfer function is then defined as the ratio between the cross-power spectrum of the cleaned mock with the input mock and the auto-power spectrum of the input mock,
\begin{equation}
    \mathcal{T}(k) = \Bigl\langle \frac{\mathcal{P}[m_{\mathrm{mock}}, m_{\mathrm{clean,mock}}]}{\mathcal{P}[m_{\mathrm{mock}}, m_{\mathrm{mock}}]} \Bigr\rangle_{\mathrm{m}},
    \label{eq:transfer_function}
\end{equation}
where $\mathcal{P}[\,,]$ denotes the cross-power spectrum of two fields and $\langle\rangle_{\mathrm{m}}$ the average over mock realisations. We use $100$ realisations, which we have verified to be sufficient for the mean to be stable, and the measured cross-power spectrum is corrected as $P_\mathrm{H\textsc{i},g}(k)\to P_\mathrm{H\textsc{i},g}(k)/\mathcal{T}(k)$. We have also run $100$ null realisations, in which the mock is injected but the cross-correlation is taken with an uncorrelated galaxy field, and verified that the recovered power is consistent with zero in every bin, confirming that the injection procedure does not introduce a spurious signal. We estimate the uncertainty on the measured monopole analytically, under the assumption of Gaussian statistics,
\begin{equation}
    \sigma^2_\mathrm{H\textsc{i},g}(k) = \frac{1}{2N_{\mathrm{modes}}(k)}\Bigl[P^2_\mathrm{H\textsc{i},g}(k)/\mathcal{T}^2(k) + P_{\mathrm{\hi}}(k)\,P_{\mathrm{g}}(k)/\mathcal{T}(k)\Bigr],
    \label{eq:gaussian_errors}
\end{equation}
where $P_{\mathrm{\hi}}$ and $P_{\mathrm{g}}$ are the measured \hi\ and galaxy auto-power spectra, the latter including the shot noise, and $N_{\mathrm{modes}}$ is the number of Fourier modes contributing to the bin after the cuts described above. We motivate the choice of the covariance estimation in \appref{apdx:cov}. The galaxy shot noise is estimated from the gridded catalogue and corrected for the effect of the assignment scheme following \citet{Sefusatti:2015aex}.

It is worth emphasising that the transfer function is a positive multiplicative rescaling, and therefore corrects the amplitude of the measurement without altering its sign or the shape of the window function. The corrected data vector, with the covariance described above, is the input to the likelihood analysis presented in \secref{sec:results}.

\subsection{Modelling the cross-correlation power spectrum}
\label{subsec:model}
We now outline the modelling of the power spectrum, following \citet{Cunnington:2022uzo, Berti:2022ilk, Casas:2022vik, Berti:2023viz, 2025MNRAS.537.3632M, Autieri:2025sxz}. The model of the cross-correlation power spectrum is given by
\begin{equation}
    P_{\mathrm{\hi,g}}(k,\mu) = \frac{\Tb\bigl(r\bHI \bg  + \bHI f\mu^2 + \bg f\mu^2 + f^2 \mu^4\bigr)P_\mathrm{m}}{1+ (k\mu\sigma_\mathrm{v}/H_0)^2},
    \label{eq:ps_model}
\end{equation}
where $\Tb$ is the mean \hi\ brightness temperature, $\bHI$ and $\bg$ are the \hi\ and galaxy biases, $r$ is the correlation coefficient between \hi\ fluctuations and galaxy overdensities, $f$ is the growth rate of structures, $\mu=\hat{k}\cdot\hat{z}$ is the cosine of the angle between the wavevector and the line of sight, and $P_\mathrm{m}$ is the matter power spectrum computed with \textsc{CAMB} \citep{Lewis:1999bs, 2026arXiv260714854L}. The denominator is intended to model the Fingers-of-God (FoG) effect, which suppresses clustering on small scales. The FoG suppression is controlled by the velocity dispersion of the tracer $\sigma_\mathrm{v}$. The cross-correlation coefficient $r$ is introduced to account for additional unknown effects that may change the estimated correlation degree between the two tracers (see e.g. \citealt{Wolz:2015ckn, Obuljen:2022cjo}). The mean \hi\ temperature $\Tb$ can be computed in terms of the \hi\ density fraction $\Omega_{\rm \hi}$ as \citep{2013MNRAS.434.1239B}
\begin{equation}
    \Tb(z)=180\,\Omega_{\rm \hi}(z) \,h\frac{(1+z)^2}{\sqrt{\Omega_\mathrm{m}(1+z)^3+\Omega_\mathrm{\Lambda}}} \,\mathrm{mK},
    \label{eq:brightness_temp}
\end{equation}
where $\Omega_\mathrm{m}$ and $\Omega_{\mathrm{\Lambda}}$ are the density parameters of matter and the cosmological constant. When fitting the data using the model, the galaxy bias is fixed to $\bg = 0.911$ \citep{2011MNRAS.415.2876B}, since it is expected to be well constrained from analyses of the galaxy auto-power spectrum. Similarly, the growth rate $f$ can be approximated as $f\approx \Omega_\mathrm{m}(z)^{\gamma}$, with $\gamma=0.545$ \citep{Linder:2005in, Planck:2018nkj}. 
To correctly model the signal, we must also include observational and mapping effects. In single-dish intensity mapping experiments, the observed density field is effectively smoothed by the telescope beam in the direction perpendicular to the line of sight. Assuming the beam can be modelled as a Gaussian, the smoothing scale of the beam is defined as $R_{\mathrm{beam}}=\sigma_{\rm \theta}\, \chi(z)$, where $\sigma_{\rm \theta} = \theta_{\rm{FWHM}}/(2\sqrt{2\ln 2})$, $\theta_{\rm{FWHM}}$ is the full width at half maximum of the beam computed in radians, and $\chi(z)$ is the comoving distance to redshift $z$. Since our redshift bin is very narrow, we can safely compute the physical beam size at the effective redshift of the data, $z_{\rm{eff}}\sim 0.425$. Then, in Fourier space, the beam damping term is 
\begin{equation}
    \tilde{B}_{\rm beam}(k,\mu) = \exp\biggl(\frac{-k^2 R_{\rm beam}^2(1-\mu^2)}{2}\biggr).
    \label{eq:beam_attenuation}
\end{equation}
Furthermore, as described above, the galaxy positions and intensity map voxels need to be regridded onto a regular 3D grid in Cartesian coordinates, and we achieve this by following the method discussed in \citet{Cunnington:2023aou, 2025MNRAS.537.3632M} and presented above. This procedure has the effect of smoothing the field, and we therefore introduce a correction for this in the power spectrum model to compensate for the gridding effect, which takes the form \citep{Sefusatti:2015aex}
\begin{equation}
    \tilde{W}_{\mathrm{ngp}}(\boldsymbol{k}) = \mathrm{sinc}\biggl(\frac{k_{\mathrm{x}}H_{\mathrm{x}}}{2}\biggr) \mathrm{sinc}\biggl(\frac{k_{\mathrm{y}}H_{\mathrm{y}}}{2}\biggr)
    \mathrm{sinc}\biggl(\frac{k_{\mathrm{z}}H_{\mathrm{z}}}{2}\biggr), 
    \label{eq:gridding_compensation}
\end{equation}
where $\mathrm{sinc}(x) = \sin(x)/x$ and $H$ is the cell size. 
Additionally, there are distortions to the power due to the finite resolution of the intensity maps. We compute the pixel size in comoving units as $s_{\mathrm{pix}}=\Delta \theta \,\chi(z_{\mathrm{eff}})$ with $\Delta \theta =0.3$ deg being the pixel size used in the intensity maps. The frequency channel depth instead is computed as the mean spacing between channel boundaries, $s_{\mathrm{chan}} = \sum_i \left(\chi(z_i) - \chi(z_{i+1})\right)/N_{\mathrm{chan}}$. With these quantities, the correction functions are computed as 
\begin{equation}
    \tilde{B}_{\mathrm{pix}}(k,\mu) = \mathrm{sinc}\biggl(\frac{k\sqrt{1-\mu^2}\,s_{\mathrm{pix}}}{2}\biggr),
    \label{eq:Bpix}
\end{equation}
and
\begin{equation}
    \tilde{B}_{\mathrm{chan}}(k,\mu) = \mathrm{sinc}\biggl(\frac{k\mu\,s_{\mathrm{chan}}}
    {2}\biggr).
    \label{eq:Bchan}
\end{equation}

Moreover, when transforming the measured redshifts into distances, one has to assume a fiducial cosmology. If the assumed cosmology does not match the true one, anisotropies arise in the measured power spectrum. This is the Alcock--Paczy\'{n}ski (AP) effect \citep{Alcock:1979mp}. These anisotropies can be modelled, in the transverse and radial directions respectively, as 
\begin{equation}
    \alpha_{\perp} (z) = \frac{D_{\rm A}}{D^{\rm fid}_{\rm A}}, \qquad \alpha_{\parallel}(z) = \frac{H^{\rm fid}(z)}{H(z)},
    \label{eq:ap_parameters}
\end{equation}
where $D_{\rm A}$ and $H(z)$ are the angular diameter distance and the Hubble parameter at redshift $z$ in the sampled cosmology, and the superscript $\rm{fid}$ denotes the fiducial values. Then, the true underlying values of $k$ and $\mu$ can be recovered as
\begin{align}
k &= \frac{k^{\rm fid}}{\alpha_\perp}\sqrt{1+(\mu^{\mathrm{fid}})^2\biggl(\frac{\alpha_\perp^2}{\alpha_\parallel^2}-1\biggr)}, \\
\mu &=\frac{\alpha_\perp \mu^{\mathrm{fid}}}{\alpha_\parallel \sqrt{1+(\mu^{\mathrm{fid}})^2\biggl(\frac{\alpha_\perp^2}{\alpha_\parallel^2}-1\biggr)}}.
\label{eq:true_k_and_mu}
\end{align}
Then the model and all the additional observational effects are discretised on the same 3D $k$-grid as the data and convolved with the window function. The observed \hi--galaxy cross-correlation power spectrum then becomes
\begin{align}
P^{\mathrm{obs}}_{\mathrm{\hi,g}}&(k^{\mathrm{fid}},\mu^{\mathrm{fid}}) = W_{\mathrm{\hi}}(\boldsymbol{k}) W^{*}_{\mathrm{g}}(\boldsymbol{k})\, \circledast  \\
    & \frac{\bigl[P_{\mathrm{\hi, g}}(k,\mu)\, \tilde{B}_{\mathrm{beam}}(k,\mu)\tilde{B}_{\mathrm{pix}}(k,\mu) \tilde{B}_{\mathrm{chan}}(k,\mu) \tilde{W}^2_{\mathrm{ngp}}(\boldsymbol{k}) \bigr]}{\alpha_\parallel \alpha_\perp^2 \sum_{\boldsymbol{x}}w_{\mathrm{\hi}}(\boldsymbol{x})w_{\mathrm{g}}(\boldsymbol{x})W_{\mathrm{\hi}}(\boldsymbol{x}) W_{\mathrm{g}}(\boldsymbol{x})},
    \label{eq:observed_pk_model}
\end{align}
where $\circledast$ denotes a convolution in $k$ space. Finally, the power spectrum model is spherically averaged into the same $k$-bins as the data.

\section{Results}
\label{sec:results}

In this section we present the constraints on cosmological parameters obtained from the \hi--galaxy cross-correlation power spectrum data. We stress that our aim is not to provide competitive constraints, but rather to demonstrate that the full pipeline, from the foreground cleaning of the intensity maps, through the power spectrum estimation, to the modelling and the likelihood evaluation, can be run end-to-end and returns unbiased and internally consistent results. Accordingly, we place the emphasis on the relative behaviour of the different runs rather than on the absolute size of the error bars.

The section is organised as follows. We begin in \secref{subsec:planck_cross} by combining the cross-correlation likelihood with the \textit{Planck} 2018 likelihoods. This run isolates the amplitude of the \hi\ signal from the cosmological information, allows us to compare our measurement of $\Omega_{\mathrm{\hi}}$ with the value reported by \citet{Cunnington:2022uzo} for the same field, and thereby validates the external input that we adopt in the rest of the analysis. We then present in \secref{subsec:main_result} our reference result, obtained from the cross-correlation data alone with all the nuisance parameters held fixed. We then aim to quantify the impact of the assumptions entering that configuration. First, we relax the assumption of the \hi\ abundance by sampling either $\Omega_{\mathrm{\hi}}$ or the combination $\Tb\sigma_8$ (\secref{subsec:omega_hi}). We then include the FoG damping and marginalise over the velocity dispersion $\sigma_\mathrm{v}$ (\secref{subsec:fog}). Then, using synthetic data, we quantify the bias that would follow from neglecting the observational effects (\secref{subsec:obs_effects}). We switch off the AP effect to assess its contribution to the constraints (\secref{subsec:ap}). Finally, in \secref{sec:robustness} we test the robustness of the result against the choices made in the data analysis.

Unless stated otherwise, all the runs share the following setup. We sample $\Omega_\mathrm{c}h^2$ and $100\theta_{\mathrm{MC}}$ with uniform priors. On $\Omega_\mathrm{b}h^2$ we impose a Gaussian prior from Big Bang Nucleosynthesis (BBN), $\Omega_\mathrm{b}h^2 = 0.02218 \pm 0.00055$ \citep{Schoneberg:2024ifp, Burns:2023sgx}, which is independent of the CMB and of the late-time expansion history. We fix the remaining cosmological parameters to the \textit{Planck} 2018 best-fit values \citep{Planck:2018vyg}, namely the amplitude and tilt of the primordial power spectrum, $\ln(10^{10}A_\mathrm{s})$ and $n_\mathrm{s}$, and the optical depth to reionisation $\tau$. This choice is motivated by the fact that $n_\mathrm{s}$ and $\tau$ are tightly constrained by the CMB and $A_\mathrm{s}$ is completely degenerate with $\Omega_\mathrm{\hi}$. Furthermore, in the following we verify that our cross-correlation data do not add any information to CMB data. Derived parameters, in particular $\Omega_\mathrm{m}$, $H_0$ and $\sigma_8$, are obtained from the sampled ones. The galaxy bias is fixed to $\bg=0.911$ \citep{2011MNRAS.415.2876B}, since it can be independently determined from the galaxy auto-power spectrum, the \hi\ bias is fixed to $\bHI=1.13$, as obtained from hydrodynamical simulations at the effective redshift of our measurement \citep{Villaescusa-Navarro:2018vsg}, and the cross-correlation coefficient is fixed to $r=0.9$, consistent with the value inferred in \citet{Chang:2010jp}. We quote one-dimensional marginalised constraints as the posterior mean with $68\%$ credible intervals, and we show two-dimensional contours at the $68\%$ and $95\%$ levels. Throughout, we focus the discussion on $\Omega_\mathrm{m}$, $\Omega_\mathrm{c}h^2$, $H_0$ and $\sigma_8$.

\subsection{Combination with \textit{Planck} 2018 and the \hi\ abundance}
\label{subsec:planck_cross}

The overall amplitude of the model of \autoref{eq:ps_model} is controlled by the product $\Tb\,\bHI\,\bg\,r$ together with the amplitude of the matter power spectrum, so that the \hi\ density parameter is strongly degenerate with $\sigma_8$ and, through it, with the cosmological parameters. Before using the cross-correlation data on their own, it is therefore useful to isolate the amplitude of the \hi\ signal from the cosmological information, which can be done by combining our likelihood with CMB data.

\begin{figure}
\centering
\includegraphics[width=0.50\textwidth]{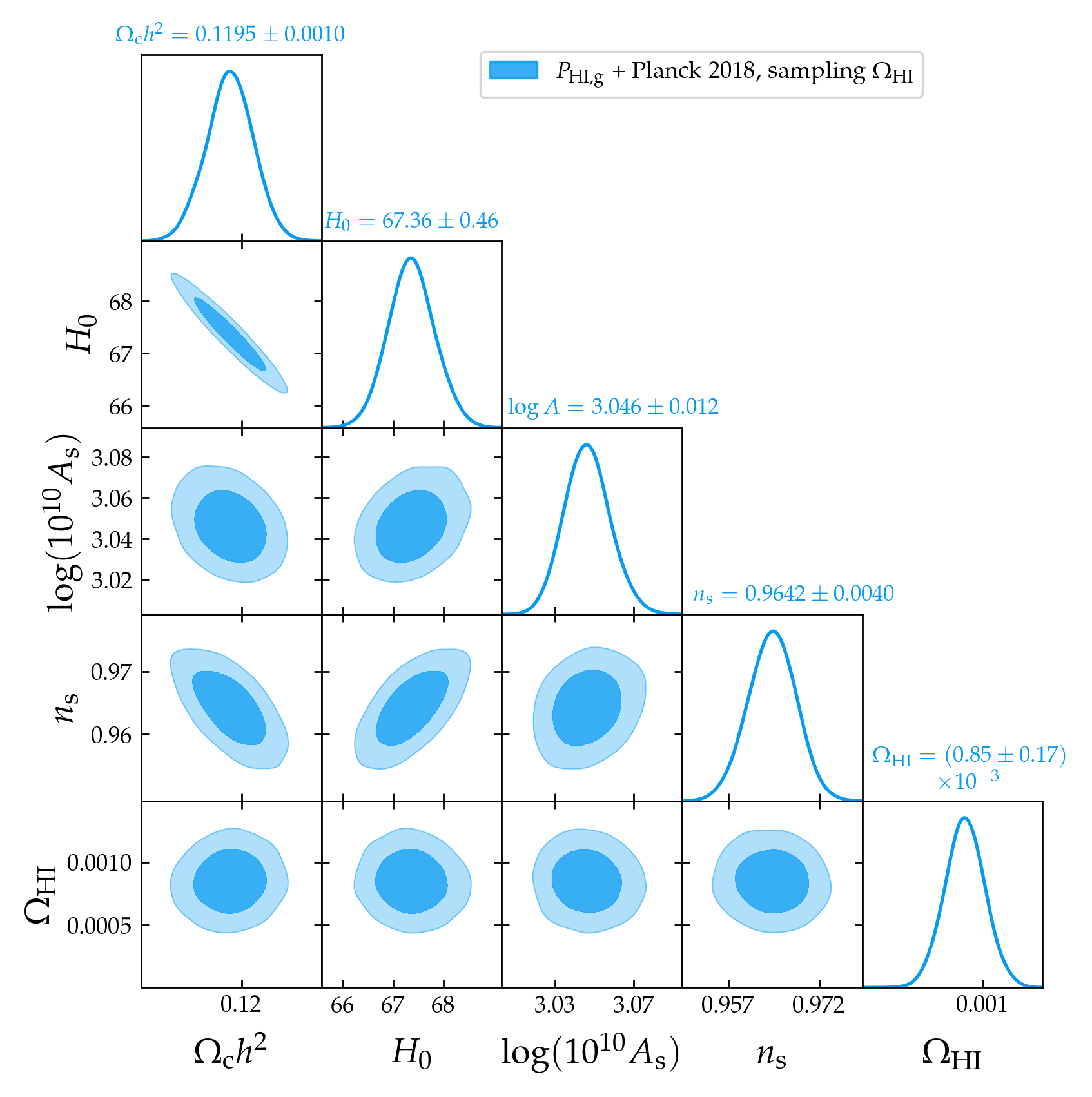}
\caption{\label{fig:cross_planck}
Constraints from the combination of the cross-correlation data with the \textit{Planck} 2018 likelihoods (including CMB lensing), with all cosmological parameters and $\Omega_{\mathrm{\hi}}$ sampled. }
\end{figure}

We begin by running the cross-correlation likelihood in combination with the \textit{Planck} 2018 TT,TE,EE+lowE+lensing likelihoods \citep{Planck:2019nip}, varying all six $\Lambda$CDM parameters and sampling $\Omega_{\mathrm{\hi}}$ with a wide uniform prior. In this configuration, the cosmological parameters are entirely determined by the CMB, and the cross-correlation data constrain the amplitude of the \hi\ signal alone. The resulting posteriors are shown in \autoref{fig:cross_planck}: the constraints on the cosmological parameters are unchanged with respect to those obtained from \textit{Planck} alone, as expected, given the relative statistical weight of the two data sets, while $\Omega_{\mathrm{\hi}}$ is measured at the $20\%$ level, $\Omega_{\mathrm{\hi}}=(0.85\pm0.17)\times10^{-3}$. This value is consistent within $1\sigma$ with the constraint on $\Omega_{\mathrm{\hi}}b_{\hi}r$ reported by \citet{Cunnington:2022uzo} for the same field, once the same values of $\bHI$ and $r$ are assumed. This agreement is reassuring: it shows that the amplitude of the cross-power spectrum recovered by our pipeline is consistent with previous analyses of the same data performed with independent estimators and modelling choices. Moreover, it supports the assumption that a determination of $\Omega_{\mathrm{\hi}}$ at this level of precision from independent observations is a realistic external input, which is what motivates holding it fixed in the reference analysis presented next.

\subsection{Reference result}
\label{subsec:main_result}

\begin{table}
\centering
\caption{Configuration of the reference analysis and resulting constraints. Sampled parameters are listed with their priors, where $\mathcal{U}[a,b]$ denotes a uniform distribution and $\mathcal{N}(\mu,\sigma)$ a Gaussian with mean $\mu$ and standard deviation $\sigma$. $\sigma_8$, $H_0$ and $\Omega_\mathrm{m}$ are derived from the sampled parameters. Constraints are posterior means with $68\%$ credible intervals. The lower block lists the parameters held fixed in the reference analysis and their fiducial values.}
\label{tab:reference}
\begin{tabular}{lcc}
\hline
Parameter & Prior & Reference result \\
\hline
$\Omega_\mathrm{c}h^2$ & $\mathcal{U}[0.05,\,0.99]$ & $0.146^{+0.035}_{-0.070}$ \\[2pt]
$\Omega_\mathrm{b}h^2$ & $\mathcal{N}(0.02218,\,0.00055)$ & $0.02218\pm0.00055$ \\[2pt]
$100\theta_{\mathrm{MC}}$ & $\mathcal{U}[0.85, \,1.2]$ & $1.050\pm 0.025$ \\[2pt]
$H_0\;[\mathrm{km\,s^{-1}\,Mpc^{-1}}]$ & derived & $65.3^{+5.8}_{-10}$ \\[2pt]
$\sigma_8$ & derived & $0.86^{+0.19}_{-0.12}$ \\[2pt]
$\Omega_\mathrm{m}$ & derived & $0.44^{+0.12}_{-0.30}$ \\
\hline
\multicolumn{3}{l}{\textit{Fixed parameters}} \\
\hline
$\ln(10^{10}A_\mathrm{s})$ & \multicolumn{2}{c}{$3.0448$} \\
$n_\mathrm{s}$ & \multicolumn{2}{c}{$0.96605$} \\
$\tau$ & \multicolumn{2}{c}{$0.0561$} \\
$\Omega_{\mathrm{\hi}}$ & \multicolumn{2}{c}{$0.83\times10^{-3}$} \\
$\bHI$ & \multicolumn{2}{c}{$1.13$} \\
$\bg$ & \multicolumn{2}{c}{$0.911$} \\
$r$ & \multicolumn{2}{c}{$0.9$} \\
$\sigma_\mathrm{v}\;[\mathrm{km\,s^{-1}}]$ & \multicolumn{2}{c}{$0$} \\
\hline
\end{tabular}
\end{table}

\begin{figure}
\centering
\includegraphics[width=0.5\textwidth]{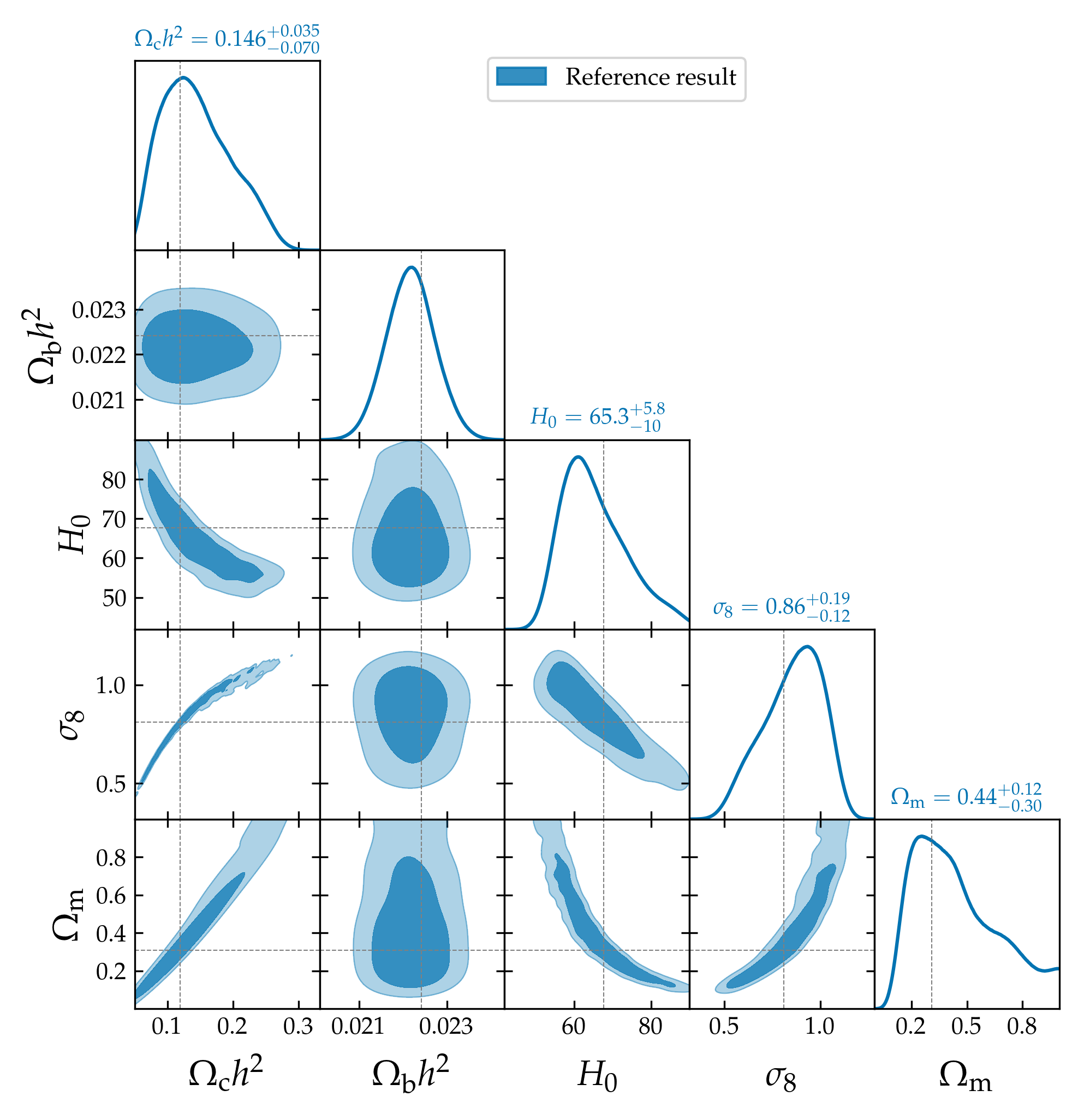}
\caption{\label{fig:reference_result}
Triangle plot of the reference result, showing $\Omega_\mathrm{c}h^2$, $\Omega_\mathrm{b}h^2$, $H_0$, $\sigma_8$ and $\Omega_\mathrm{m}$. Dashed lines indicate the \textit{Planck} 2018 cosmology.}
\end{figure}

We now define the configuration that we adopt as our reference. All the nuisance parameters entering \autoref{eq:ps_model} are held fixed: $\bg=0.911$, $\bHI=1.13$, $r=0.9$ and $\Omega_{\mathrm{\hi}}=0.83\times10^{-3}$. The FoG damping is switched off, $\sigma_\mathrm{v}=0$, and the growth rate is computed from the sampled cosmology as $f=\Omega^{\gamma}_\mathrm{m}(z)$. We have also repeated the analysis with $f$ fixed to $f=0.737$, and found constraints very similar to those obtained when $f$ is computed from the sampled cosmology. We adopt $f=\Omega_{\rm m}^\gamma$ in the reference analysis and in all subsequent tests.
The sampled parameters are therefore $\Omega_\mathrm{c}h^2$, $\Omega_\mathrm{b}h^2$ and $100\theta_{\mathrm{MC}}$, with the BBN prior applied to $\Omega_\mathrm{b}h^2$, and $\Omega_\mathrm{m}$, $H_0$ and $\sigma_8$ are derived. Each of these choices is supported by an external measurement or, as shown in the following subsections, has been verified not to bias the result. 

The resulting posteriors are shown in \autoref{fig:reference_result}. We obtain $\Omega_\mathrm{c}h^2=0.146^{+0.035}_{-0.070}$, $H_0=65.3^{+5.8}_{-10}\;\mathrm{km\,s^{-1}\,Mpc^{-1}}$ and $\Omega_\mathrm{m}=0.44^{+0.12}_{-0.30}$. All three are consistent within $1\sigma$ with the \textit{Planck} 2018 values, which are indicated by the dashed lines in the figure. The posterior on $\Omega_\mathrm{b}h^2$ simply reproduces the BBN prior, as expected, confirming that the cross-correlation data carry no independent information on the baryon density. The $\Omega_\mathrm{c}h^2$--$H_0$ and $\Omega_\mathrm{m}$--$\sigma_8$ planes display the degeneracy directions characteristic of a clustering measurement of fixed amplitude, and the $\Omega_\mathrm{m}$ posterior is visibly non-Gaussian, with a skewed distribution towards low values which is responsible for the asymmetry of the quoted interval.

We emphasise how these numbers should be interpreted. They are obtained under the assumption that all the astrophysical nuisance parameters are known exactly, which is the most favourable situation that can realistically be envisaged for this kind of measurement. The tests presented in \secref{subsec:omega_hi} and \secref{subsec:fog} show explicitly how the constraints relax when this assumption is weakened. 

\subsection{Sampling the amplitude of the signal}
\label{subsec:omega_hi}

\begin{figure}
\centering
\includegraphics[width=0.5\textwidth]{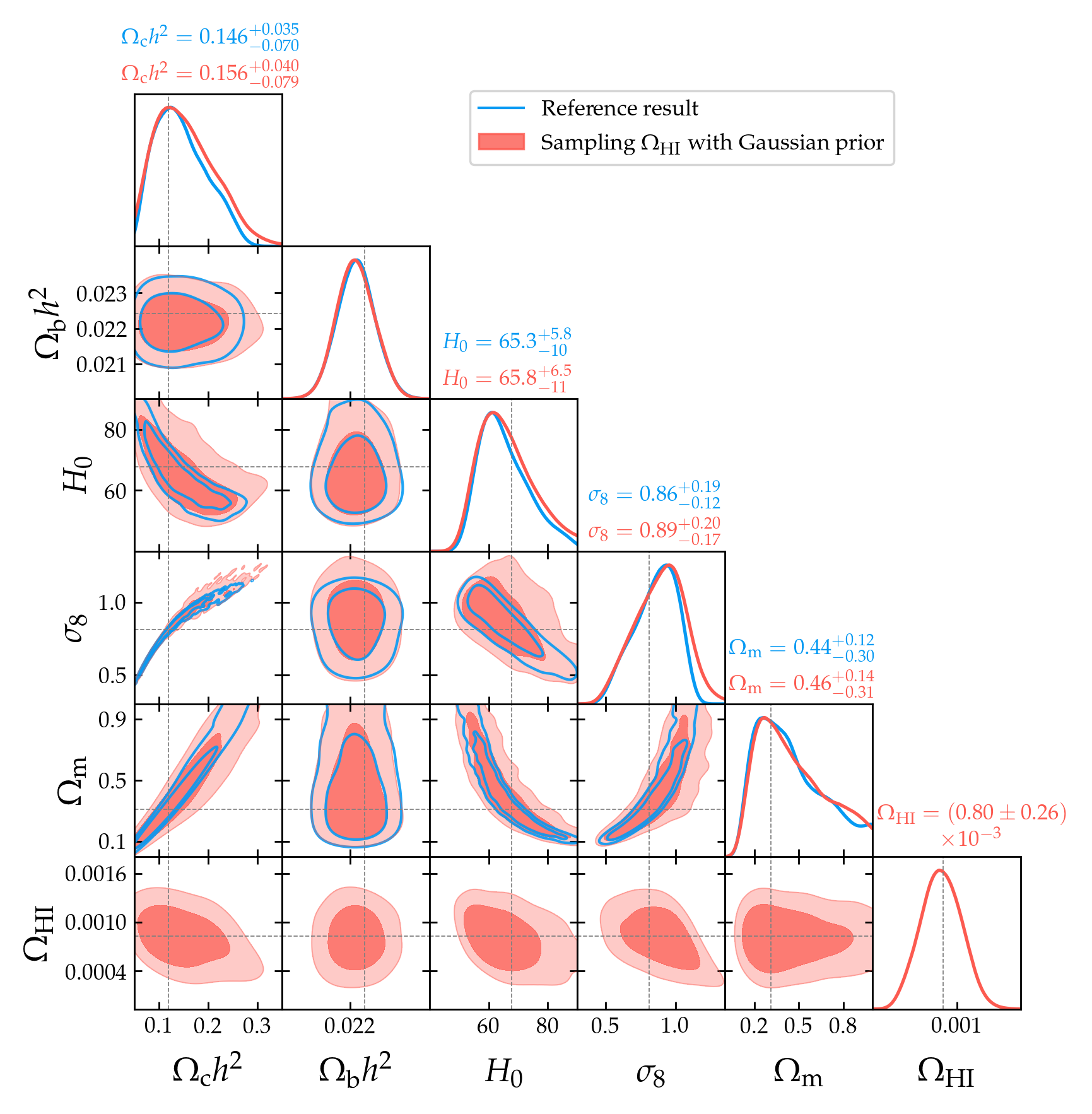}
\caption{\label{fig:omegaHI_gauss}
Marginalised posteriors for the reference run (blue) and for the run in which $\Omega_{\mathrm{\hi}}$ is sampled with a Gaussian prior centred on the fiducial value (red). Dashed lines indicate the \textit{Planck} 2018 best-fit values.}
\end{figure}

\begin{figure}
\centering
\includegraphics[width=0.5\textwidth]{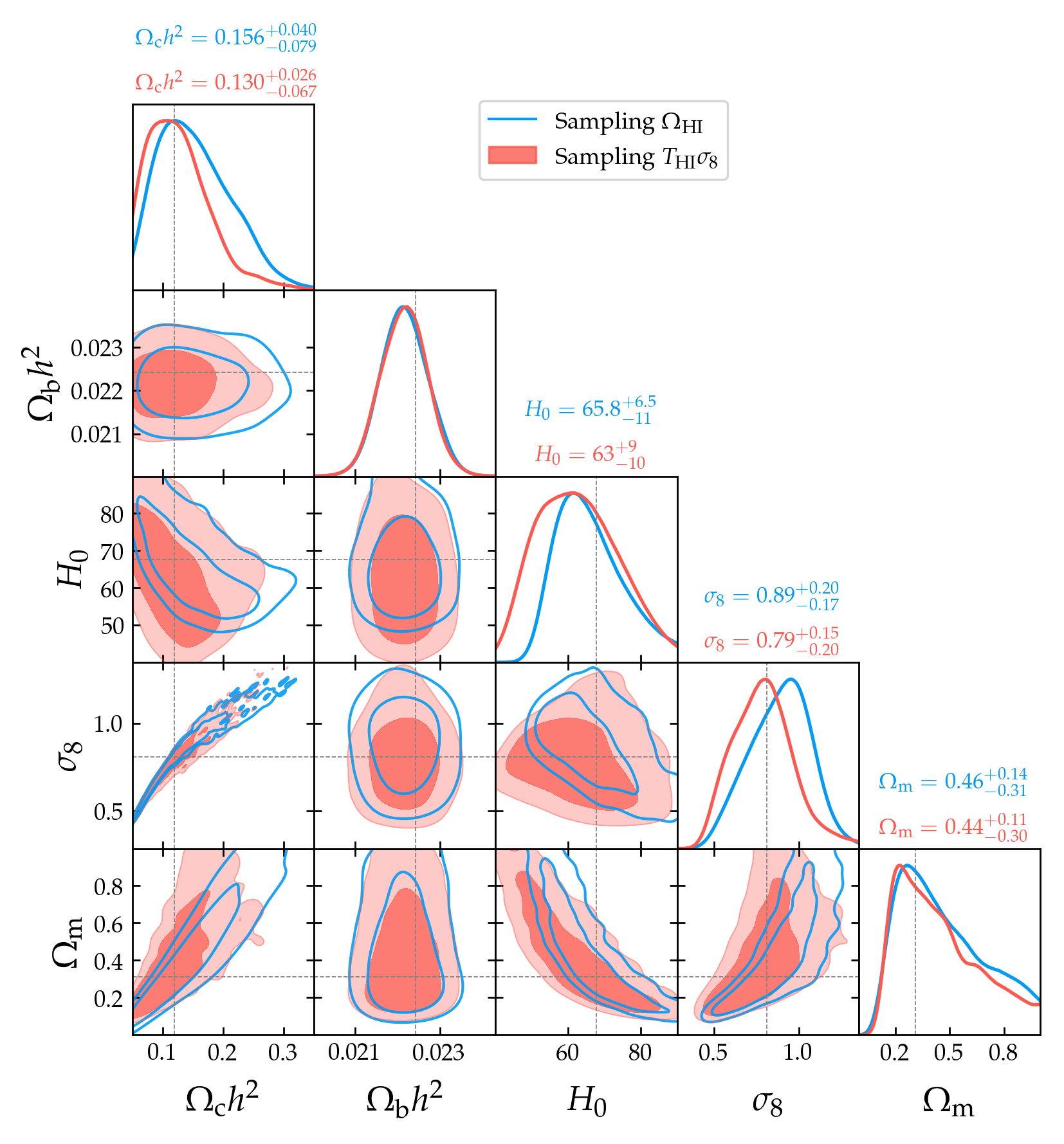}
\caption{\label{fig:Tbsigma8_vs_omegaHI}
Comparison between sampling $\Omega_{\mathrm{\hi}}$ (blue) and sampling the combination $\Tb\sigma_8$ (red), in both cases with a Gaussian prior of the same relative width. The two parametrisations return consistent constraints on the cosmological parameters.}
\end{figure}

We now relax the assumption that the \hi\ abundance is known and quantify how much of the constraining power of the measurement is lost. As already noted, $\Omega_{\mathrm{\hi}}$ controls the overall amplitude of \autoref{eq:ps_model} and is therefore strongly degenerate with the amplitude of the matter power spectrum and, through it, with the other cosmological parameters. Since $\Tb$ also depends explicitly on $\Omega_\mathrm{m}$ and $h$ through \autoref{eq:brightness_temp}, this degeneracy is not simply a rescaling, and it propagates into the geometry of the parameter space. It is therefore important to establish how much of the constraining power of the measurement comes from the shape of the power spectrum and how much from its amplitude.

We first consider the case in which $\Omega_{\mathrm{\hi}}$ is sampled, together with the cosmological parameters, with a Gaussian prior that follows the result of \citet{Cunnington:2022uzo}, $\Omega_{\mathrm{\hi}}=(0.83\pm0.26)\times10^{-3}$. This case is shown in \autoref{fig:omegaHI_gauss}, compared to the reference run. The constraints show a very small degradation: the constraint on $\Omega_\mathrm{c}h^2$ degrades from $0.146^{+0.035}_{-0.070}$ to $0.156^{+0.040}_{-0.079}$, and that on $H_0$ from $65.3^{+5.8}_{-10}$ to $65.8^{+6.5}_{-11}$ $\mathrm{km\,s^{-1}\,Mpc^{-1}}$. Similarly, the posteriors of $\Omega_\mathrm{m}$ and $\sigma_8$ remain centred at comparable values and broaden from $0.44^{+0.12}_{-0.30}$ to $0.46^{+0.14}_{-0.31}$ and from $0.86^{+0.19}_{-0.12}$ to $0.89^{+0.20}_{-0.17}$, respectively. The comparison quantifies in a transparent way the cost of not knowing the \hi\ abundance: with a prior at the $\sim 31\%$ level on $\Omega_{\mathrm{\hi}}$, the loss in constraining power on $\Omega_\mathrm{m}$ is close to none, whereas with no prior at all the measurement is no longer informative on the cosmological parameters.

Rather than $\Omega_{\mathrm{\hi}}$, one can instead directly sample the combination $\Tb\sigma_8$, which is the quantity to which the monopole is most sensitive. This has the advantage of not imposing the explicit $\Omega_\mathrm{m}$ and $h$ dependence of \autoref{eq:brightness_temp} on a parameter that is in practice unknown. \autoref{fig:Tbsigma8_vs_omegaHI} compares the two parametrisations, in both cases with a Gaussian prior of the same relative width. The resulting constraints are consistent between the two parametrisations. Sampling $\Tb\sigma_8$ yields a somewhat tighter and lower constraint on $\Omega_\mathrm{c}h^2$, $0.130^{+0.026}_{-0.067}$ compared to $0.156^{+0.040}_{-0.079}$ when $\Omega_{\mathrm{\hi}}$ is sampled directly, and a similar trend is found for $\sigma_8$, for which we find $0.79^{+0.15}_{-0.20}$ compared to $0.89^{+0.20}_{-0.17}$. In contrast, we find that sampling $\Tb\sigma_8$ results in a slightly broader constraint on $H_0$, $63^{+9}_{-10}$ compared to $65.8^{+6.5}_{-11}$\,$\rm km\,s^{-1}\,Mpc^{-1}$ when $\Omega_{\mathrm{\hi}}$ is sampled. Overall, the peaks of the posteriors agree within $1\sigma$, and we take this as an indication that the results are not strongly driven by the specific way in which the amplitude is parametrised. The posterior on $\sigma_8$ is the most affected: its mean moves from $0.89$ to $0.79$ when $\Tb\sigma_8$ is sampled, a shift of about $0.5\sigma$. We interpret this as a projection effect rather than as a different preference of the data. The two runs share the same information on the shape of the power spectrum, as shown by the nearly identical posteriors on $\Omega_\mathrm{m}$, $0.46^{+0.14}_{-0.31}$ and $0.44^{+0.11}_{-0.30}$, and differ in how the measured amplitude is divided between the \hi\ amplitude parameter and $\sigma_8$. A Gaussian prior on $\Omega_{\mathrm{\hi}}$ and a Gaussian prior on $\Tb\sigma_8$ of the same relative width are not equivalent priors on the full parameter space. As can be seen from \autoref{eq:brightness_temp}, the transformation between the two depends on $h$, on $\Omega_\mathrm{m}$ and, in the second case, on $\sigma_8$ itself. Since the data do not constrain the amplitude of the \hi\ signal independently of $\sigma_8$, the marginalised posterior on $\sigma_8$ reflects this difference in prior volume. 
It is worth stressing that, in both parametrisations, a prior on the amplitude of the \hi\ signal is sufficient to retain most of the constraining power of the reference analysis, in which the amplitude is held fixed. 
In the following, we adopt $\Omega_{\mathrm{\hi}}$ as the amplitude parameter, for ease of comparison with the literature.

\subsection{Including the Fingers-of-God damping}
\label{subsec:fog}

\begin{figure}
\centering
\includegraphics[width=0.5\textwidth]{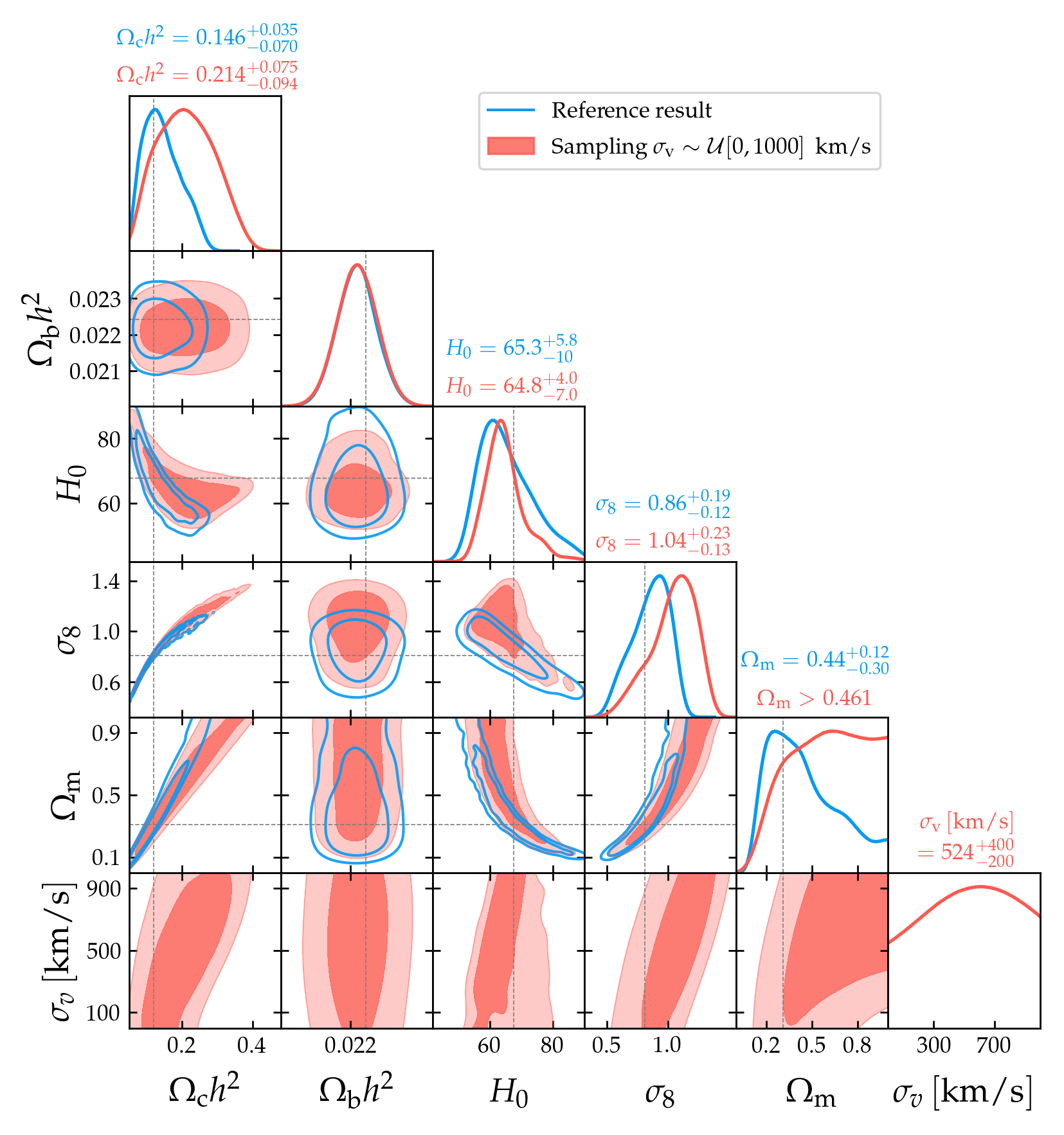}
\caption{\label{fig:fog_uniform}
Reference run (blue) compared with the run including the FoG damping and sampling $\sigma_\mathrm{v}$ with a uniform prior, $\sigma_\mathrm{v}\sim \mathcal{U}[0, 1000]\,\mathrm{km\,s^{-1}}$ (red). The velocity dispersion is largely unconstrained by the data and its marginalisation degrades the constraints on all the cosmological parameters.}
\end{figure}

\begin{figure}
\centering
\includegraphics[width=0.5\textwidth]{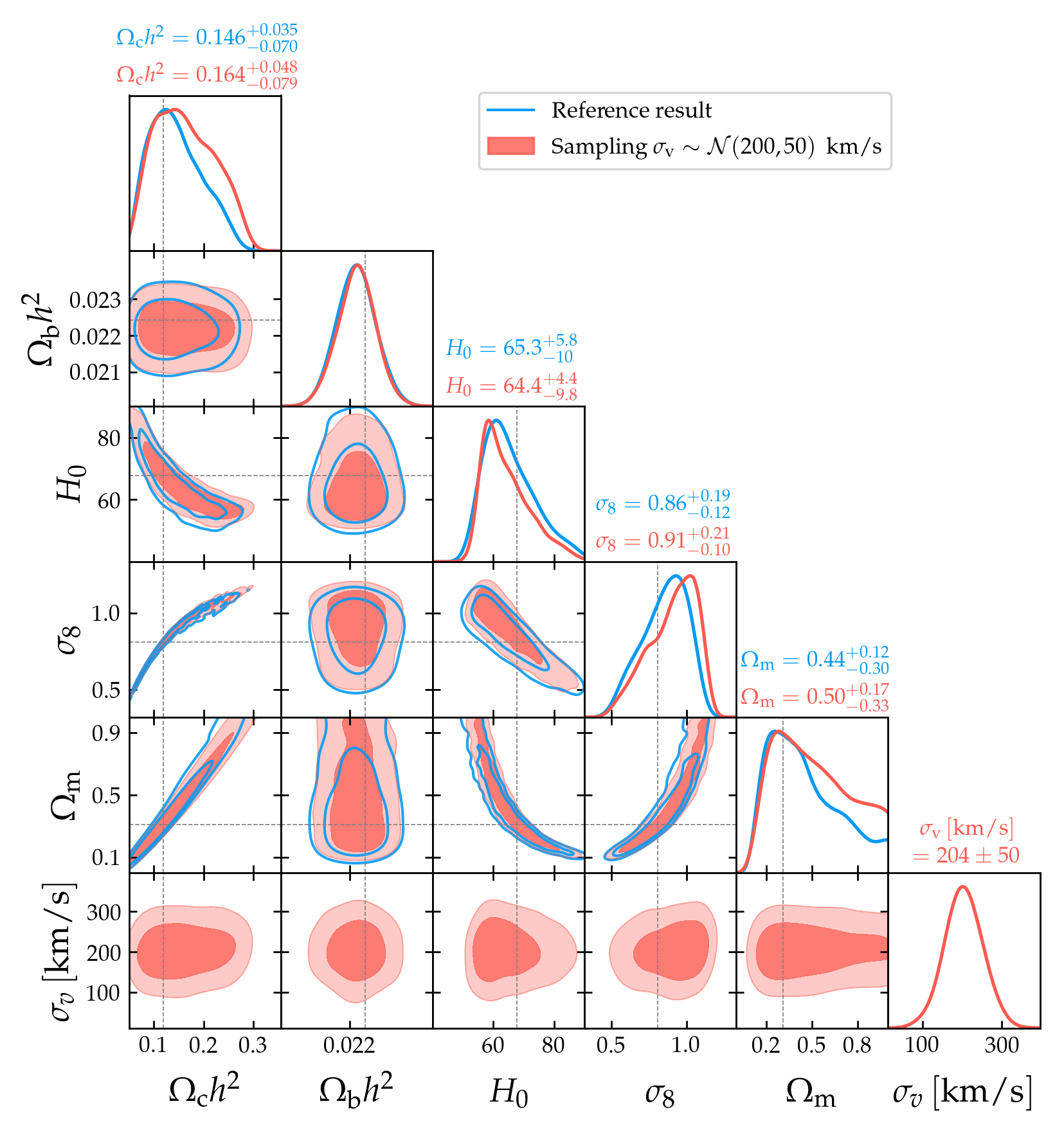}
\caption{\label{fig:fog_gauss}
Reference run (blue) compared with a run including the FoG damping and sampling $\sigma_\mathrm{v}$ with a Gaussian prior centred at $\sigma_\mathrm{v}=200\;\mathrm{km\,s^{-1}}$ with width equal to $50\;\mathrm{km\,s^{-1}}$, motivated by the analysis of the hydrodynamical simulations of \citet{Villaescusa-Navarro:2018vsg}.
}
\end{figure}

The FoG term in \autoref{eq:ps_model} suppresses power along the line of sight on small scales and is controlled by the velocity dispersion $\sigma_\mathrm{v}$. In the reference case we switch this term off by setting $\sigma_\mathrm{v}=0$. This choice is justified a posteriori by the scale cuts adopted in the analysis. For $\sigma_\mathrm{v}\simeq200\;\mathrm{km\,s^{-1}}$, representative of the values found in hydrodynamical simulations \citep{Villaescusa-Navarro:2018vsg}, the damping of purely radial modes ($\mu=1$) at the smallest scale included in the fit, $k\simeq0.3\;h\,\mathrm{Mpc^{-1}}$, is $\sim25\%$. Since the FoG term leaves transverse modes unaffected, its impact on the monopole is considerably smaller. Including all the observational effects of \autoref{eq:observed_pk_model}, the suppression is below $5\%$ for $k<0.2\;h\,\mathrm{Mpc^{-1}}$ and reaches $\sim15\%$ only in the two highest-$k$ bins, below the statistical uncertainty of these bins.

We first sample $\sigma_\mathrm{v}$ with a wide uniform prior, $\sigma_\mathrm{v}\sim \mathcal{U}[0,1000]\;\mathrm{km\,s^{-1}}$. The result is shown in \autoref{fig:fog_uniform}. As anticipated, $\sigma_\mathrm{v}$ remains essentially unconstrained: its marginalised posterior is flat over most of the prior range, with only a mild disfavouring of the smallest values. The consequence for the cosmological parameters is a substantial degradation, driven by the degeneracy between the small-scale suppression and the overall amplitude. The interval on $\Omega_\mathrm{c}h^2$ broadens from $0.146^{+0.035}_{-0.070}$ to $0.214^{+0.075}_{-0.094}$, resulting in an increase of about $60\%$ in the total width of the $1\sigma$ interval and a shift of the peak of the posterior towards a higher value. The posterior on $\Omega_\mathrm{m}$ is affected the most. It develops an extended plateau towards high values so that only a lower bound $\Omega_\mathrm{m} > 0.461$ can be placed. The constraint on $H_0$, by contrast, is only mildly affected, yielding $64.8^{+4.0}_{-7.0}$ compared to $65.3^{+5.8}_{-10}\,\mathrm{km}\,\mathrm{s}^{-1}\,\mathrm{Mpc}^{-1}$ in the reference run. We stress that the shift in $\Omega_\mathrm{c}h^2$ remains within $1\sigma$ of the reference value, so that the two results are consistent.  

Given that the data cannot constrain $\sigma_\mathrm{v}$, we have also considered a Gaussian prior centred at $\sigma_\mathrm{v}=200\;\mathrm{km\,s^{-1}}$ with a width of $50\;\mathrm{km\,s^{-1}}$, representative of the \hi\ velocity dispersion measured in the hydrodynamical simulations of \citet{Villaescusa-Navarro:2018vsg} at the redshift of our measurement. The comparison is shown in \autoref{fig:fog_gauss}. The prior recovers most of the constraining power lost in the uniform-prior case. The constraint on $\Omega_\mathrm{c}h^2$ broadens only mildly with respect to the reference run, from $0.146^{+0.035}_{-0.070}$ to $0.164^{+0.048}_{-0.079}$, and $H_0$ is essentially unaffected, $64.4^{+4.4}_{-9.8}$ compared to $65.3^{+5.8}_{-10}\;\mathrm{km\,s^{-1}\,Mpc^{-1}}$ in the reference run. The posterior on $\sigma_\mathrm{v}$ itself, $204\pm50\;\mathrm{km\,s^{-1}}$, is essentially unchanged from the input prior, indicating that the data carry no independent information on the velocity dispersion at this level of precision. The residual shift towards higher $\sigma_8$ and $\Omega_\mathrm{m}$ visible in \autoref{fig:fog_gauss}, from $0.86^{+0.19}_{-0.12}$ to $0.91^{+0.21}_{-0.10}$ and from $0.44^{+0.12}_{-0.30}$ to $0.50^{+0.17}_{-0.33}$ respectively, is the expected compensation of the small-scale suppression by an increase of the amplitude, and remains well within $1\sigma$ of the reference run. We conclude that, at the sensitivity of the present data and with the scale cuts adopted here, the treatment of the FoG is not a limiting systematic, and that setting $\sigma_\mathrm{v}=0$ in the reference case is a defensible simplification rather than an optimistic one.

This will not remain true for future measurements. With higher precision and the extension of the analysis to smaller scales, the phenomenological description adopted here, a linear Kaiser model with a single-parameter Lorentzian damping, will no longer be adequate. Perturbative approaches based on the effective field theory of large-scale structure \citep{Baumann:2010tm, Carrasco:2012cv} provide a consistent framework for this regime. This framework is now routinely used in the analysis of galaxy surveys \citep{Ivanov:2019pdj, DAmico:2019fhj}, and its application to \hi\ has started to be explored \citep{Obuljen:2022cjo}. Adapting it to the \hi--galaxy cross-correlation, and determining the range of scales over which it remains reliable for intensity mapping data, will be a necessary step for future analyses.

\subsection{Impact of observational effects in the model}
\label{subsec:obs_effects}

\begin{figure}
\centering
\includegraphics[width=0.5\textwidth]{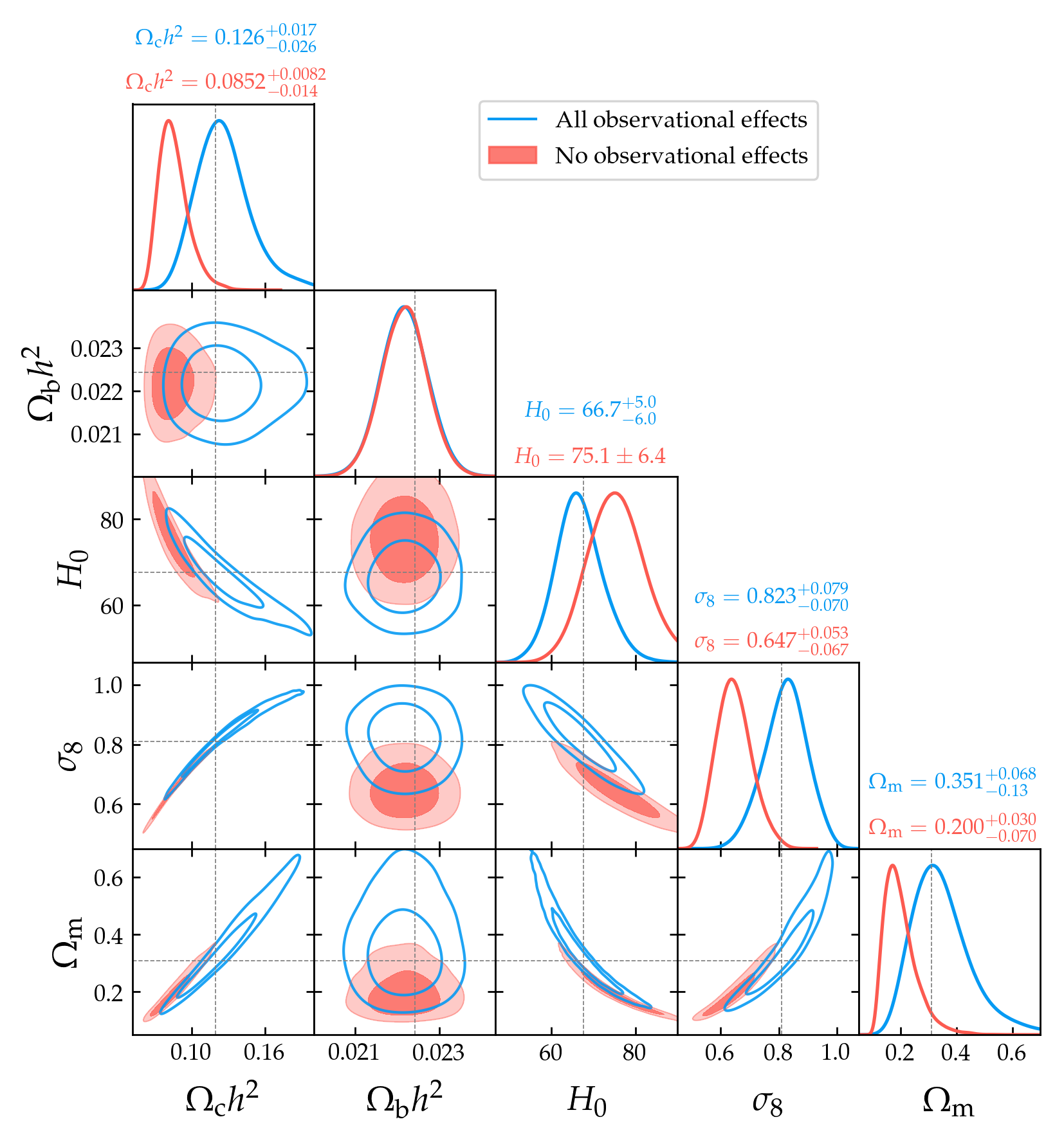}
\caption{\label{fig:obs_none}
Constraints from synthetic data generated with the full model of \autoref{eq:observed_pk_model} and analysed including all the observational effects (blue) or none of them (red). The dashed lines indicate the input cosmology used to generate the data.}
\end{figure}

\begin{figure}
\centering
\includegraphics[width=0.5\textwidth]{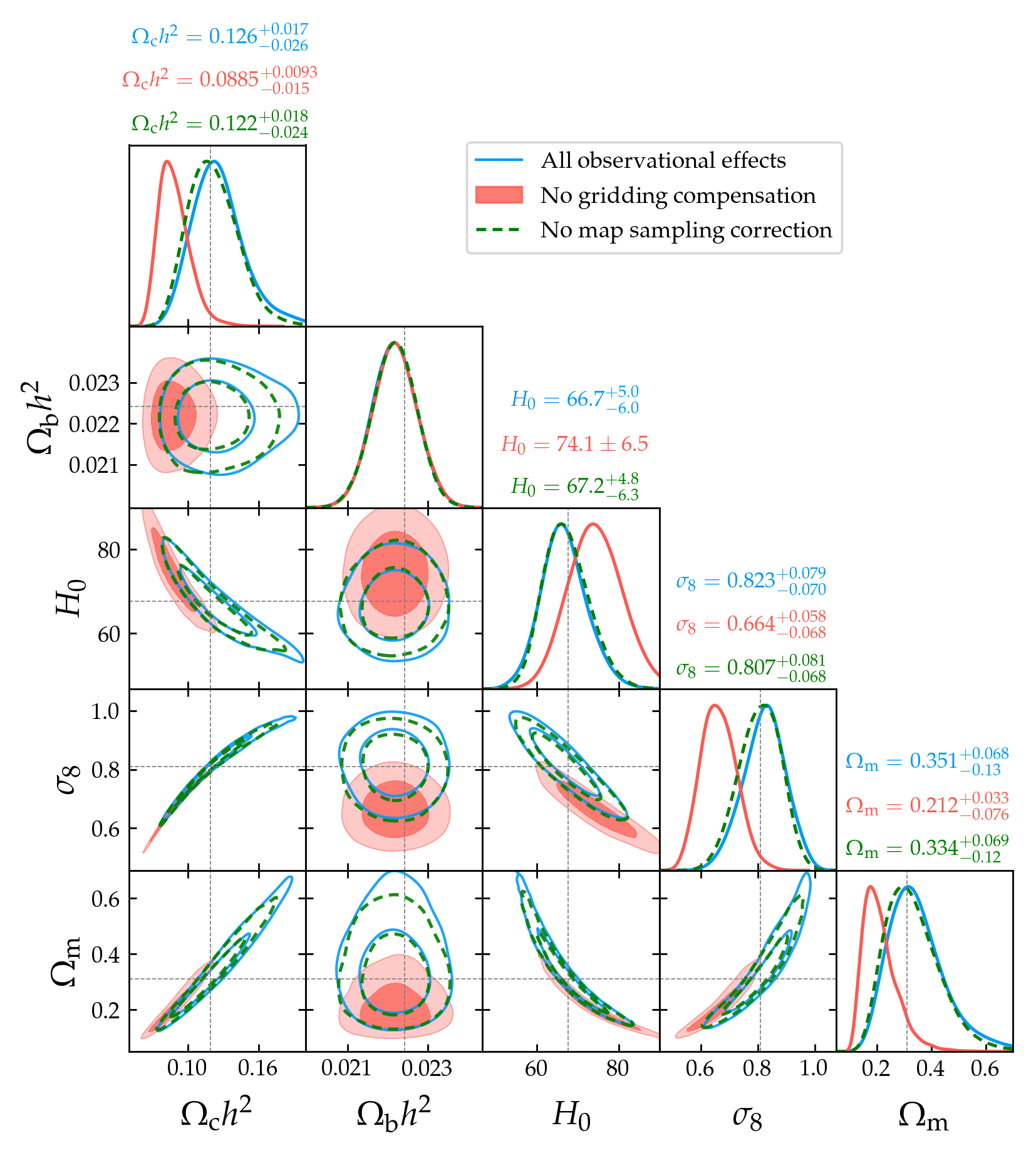}
\caption{\label{fig:obs_multiple_pieces}
As \autoref{fig:obs_none}, but comparing the analysis including all the observational effects (blue) with those in which only the gridding compensation of \autoref{eq:gridding_compensation} (red) or only the map sampling correction of \autoref{eq:Bpix} and \autoref{eq:Bchan} (green, dashed) is omitted from the model.}
\end{figure}

The modelling of the observational and mapping effects described in \secref{subsec:model} is an essential part of the pipeline.
It consists of the beam attenuation, the pixel and channel window functions, the gridding compensation and the convolution with the survey window and weights. 
To quantify the bias that would follow from omitting them, we perform a set of tests on synthetic data. 
The mock data vector is generated using the model of \autoref{eq:observed_pk_model} at a fixed reference cosmology, with the same $k$-binning, scale cuts and covariance as the real measurement. 
By construction, an analysis that includes all the effects must recover the input cosmology; any shift observed when a term is removed is therefore an unambiguous measure of the bias associated with that term, uncontaminated by the statistical fluctuations and by the modelling uncertainties of the real data.

We first consider the extreme case in which the map sampling correction, the gridding compensation and the convolution with the weights and the window function are all removed from the model, the beam being the only observational term retained. The result is shown in \autoref{fig:obs_none}. The analysis including all the effects recovers the input cosmology within $1\sigma$ for all the parameters we consider, with $\Omega_\mathrm{c}h^2=0.126^{+0.017}_{-0.026}$, $H_0=66.7^{+5.0}_{-6.0}\;\mathrm{km\,s^{-1}\,Mpc^{-1}}$, $\Omega_\mathrm{m}=0.351^{+0.068}_{-0.13}$ and $\sigma_8=0.823^{+0.079}_{-0.070}$. When the observational effects are neglected, the posteriors are instead displaced well away from the input values: we find $\Omega_\mathrm{c}h^2=0.0852^{+0.0082}_{-0.0140}$, $H_0=75.1\pm6.4\;\mathrm{km\,s^{-1}\,Mpc^{-1}}$, $\Omega_\mathrm{m}=0.200^{+0.030}_{-0.070}$ and $\sigma_8=0.647^{+0.053}_{-0.067}$, with $H_0$ biased high by $1.1\sigma$ and the others biased low by $2.2\sigma$, $1.6\sigma$ and $2.5\sigma$ respectively. 
The biased results gain a spurious increase in constraining power
for $\Omega_\mathrm{c}h^2$, $\Omega_\mathrm{m}$ and $\sigma_8$ compared to the correct ones, so that neglecting these effects yields results that appear more precise while being inaccurate. The direction of the shift can be understood by noting that neglecting the smoothing terms means attributing the observed suppression of power at high $k$ to the underlying matter power spectrum. This effect in the fit is accommodated by lowering $\Omega_\mathrm{m}h$, moving the turnover to larger scales and reducing the small-scale power relative to the large-scale one, and by reducing the overall amplitude. This test confirms that, at the sensitivity of the current data, the observational effects are not a negligible correction and cannot be ignored in the analysis.

We then isolate the individual contributions, as shown in \autoref{fig:obs_multiple_pieces}. Omitting only the gridding compensation of \autoref{eq:gridding_compensation}, while retaining all the other terms, produces a bias almost as large as that of the fully uncorrected case. We find $\Omega_\mathrm{c}h^2=0.089^{+0.009}_{-0.015}$, $\Omega_\mathrm{m}=0.212^{+0.033}_{-0.076}$, $H_0=74.1\pm6.5\;\mathrm{km\,s^{-1}\,Mpc^{-1}}$ and $\sigma_8=0.644^{+0.058}_{-0.068}$, close to those found when all effects are neglected. By contrast, omitting the map sampling correction has a negligible effect. The posteriors are almost indistinguishable from those of the full analysis, with $\Omega_\mathrm{c}h^2=0.122^{+0.018}_{-0.024}$, $H_0=67.2^{+4.8}_{-6.3}\;\mathrm{km\,s^{-1}\,Mpc^{-1}}$, $\Omega_\mathrm{m}=0.334^{+0.069}_{-0.12}$ and $\sigma_8=0.807^{+0.081}_{-0.068}$. 

The negligible impact of the map sampling correction compared with the gridding compensation follows from the different smoothing scales involved: the pixels and frequency channels of the maps are finer than the cells of the Cartesian grid, so that their window functions are much closer to unity over the range of scales used in the analysis. In the radial direction the channel width corresponds to a comoving scale below $1\;h^{-1}\,\mathrm{Mpc}$, and the channel window reduces the power by only $\sim0.1\%$ even at the highest radial wavenumber we use, $k_\parallel=0.25\;h\,\mathrm{Mpc^{-1}}$. In the transverse direction the pixel size is below the beam scale, so that the modes on which $\tilde{B}_{\mathrm{pix}}$ is appreciable are already strongly suppressed by the beam attenuation of \autoref{eq:beam_attenuation}, which is included in all the runs considered here, and carry little weight in the likelihood. Although the impact is small for the present measurement, we include the correction in the reference analysis, since its relative importance is expected to grow for data sets with higher angular resolution and lower noise.

\subsection{Switching off the Alcock--Paczy\'nski effect}
\label{subsec:ap}

\begin{figure}
\centering
\includegraphics[width=0.5\textwidth]{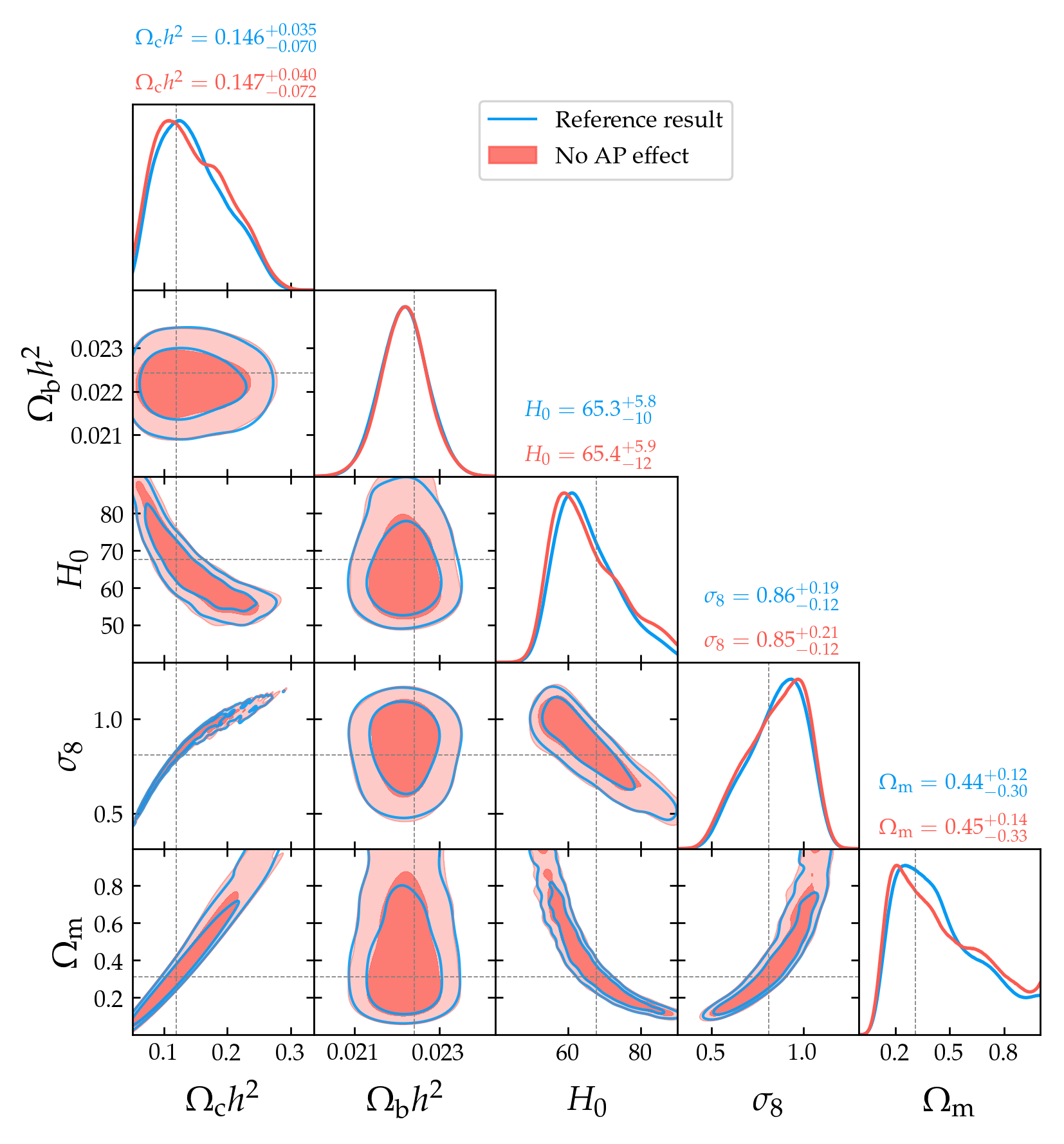}
\caption{\label{fig:ap}
Reference run (blue) compared with the run in which the AP effect is switched off in the model, i.e. $\alpha_\parallel=\alpha_\perp=1$ at every point in the chain (red). All the other settings are those of the reference configuration. Dashed lines indicate the \textit{Planck} 2018 best-fit values.}
\end{figure}

As discussed in \secref{subsec:model}, the AP distortions are anisotropies introduced into the data by the possible discrepancy between the assumed fiducial cosmology and the true underlying one. These can be modelled via the remapping of $k$ and $\mu$ in \autoref{eq:true_k_and_mu} and the volume factor $\alpha_\parallel\alpha_\perp^2$ in \autoref{eq:observed_pk_model}. 
To assess how much of the constraining power of the measurement is due to this geometric information, we repeat the reference inference setting $\alpha_\parallel=\alpha_\perp=1$ at every point in the chain, so that the model is always evaluated at the fiducial wavenumbers, independently of the sampled cosmology. The cosmological information then comes only from the shape and amplitude of $P_\mathrm{m}$ and from the growth rate. We stress that this is intended only as a test of the information content of the measurement, not as an alternative analysis choice. The AP distortions are a physical consequence of assuming a fiducial cosmology, and including them is required for a consistent model. The comparison with the reference run is shown in \autoref{fig:ap}.

The two sets of posteriors are very similar. Without the AP effect we find $\Omega_\mathrm{c}h^2=0.147^{+0.040}_{-0.072}$, $H_0=65.4^{+5.9}_{-12}\;\mathrm{km\,s^{-1}\,Mpc^{-1}}$, $\Omega_\mathrm{m}=0.45^{+0.14}_{-0.33}$ and $\sigma_8=0.85^{+0.21}_{-0.12}$, to be compared with $0.146^{+0.035}_{-0.070}$, $65.3^{+5.8}_{-10}\;\mathrm{km\,s^{-1}\,Mpc^{-1}}$, $0.44^{+0.12}_{-0.30}$ and $0.86^{+0.19}_{-0.12}$ in the reference run. The posterior means shift by less than $0.1\sigma$ for all the parameters, and the degeneracy directions in the two-dimensional planes are unchanged. The loss of constraining power is therefore small, and it follows from two features of the measurement. First, we fit only the monopole. The AP distortion can be decomposed into an isotropic dilation, $\alpha_\mathrm{iso}=(\alpha_\parallel\alpha_\perp^2)^{1/3}$, and an anisotropic distortion controlled by the ratio $\alpha_\parallel/\alpha_\perp$. To first order in $\alpha_\parallel/\alpha_\perp-1$, the anisotropic part redistributes power among different values of $\mu$ without changing its angular average, so it leaves the monopole of an isotropic spectrum unchanged and affects it only through the $\mu$ dependence of the RSD terms in \autoref{eq:ps_model}. Most of the information it carries resides in the quadrupole and higher multipoles \citep[e.g.][]{Soares:2020zaq}, which we do not use. Second, the isotropic dilation can be measured only if the power spectrum contains a feature at a known scale, namely the BAO. With the present signal-to-noise ratio, the BAO are not resolved, and over the fitted range the spectrum is effectively smooth. Therefore, the dilation acts mainly as an amplitude rescaling, adding little to no information. 

We conclude that, at the current precision, the constraints are driven mainly by the shape and amplitude of the power spectrum, and that the AP effect provides only a small additional contribution. The absence of any shift between the two runs also indicates that, at this level of precision, the results are not sensitive to the choice of the fiducial cosmology used to grid the data. Despite its small impact on the present constraints, the AP effect is a necessary part of a consistent model of the observed power spectrum, and it introduces no additional free parameters. Its role is expected to grow for the larger volumes and higher signal-to-noise ratios of future MeerKLASS surveys and of the SKAO \citep{Bull:2014rha, Berti:2023viz, Cunnington:2026opd}.

\section{Robustness against data analysis choices}
\label{sec:robustness}

The tests presented in \secref{sec:results} address the modelling side of the analysis and quantify how the inferred parameters respond to the assumptions made about the nuisance parameters and to the terms included in the theoretical power spectrum, keeping the data vector fixed. A complementary question is whether the result is stable against the choices made upstream, in the construction of the data vector itself. Two of these choices are potentially critical for an intensity mapping measurement: the number of modes removed in the PCA foreground cleaning, $N_{\mathrm{fg}}$, and the cuts applied in the $(k_\parallel, k_\perp)$ plane before the one-dimensional average. Both are, to a large extent, arbitrary, and neither is calibrated by the cosmological likelihood. In this section we repeat the reference inference on data vectors obtained with different choices of $N_{\mathrm{fg}}$ (\secref{subsec:nfg}) and of the scale cuts (\secref{subsec:scale_cuts}), and verify that the constraints are not significantly affected at the level of precision of the present measurement. Throughout this section all the modelling choices are those of the reference configuration defined in \secref{subsec:main_result}. Only the data vector and its covariance change.

\subsection{Dependence on the number of foreground modes removed}
\label{subsec:nfg}

\begin{figure}
\centering
\includegraphics[width=0.5\textwidth]{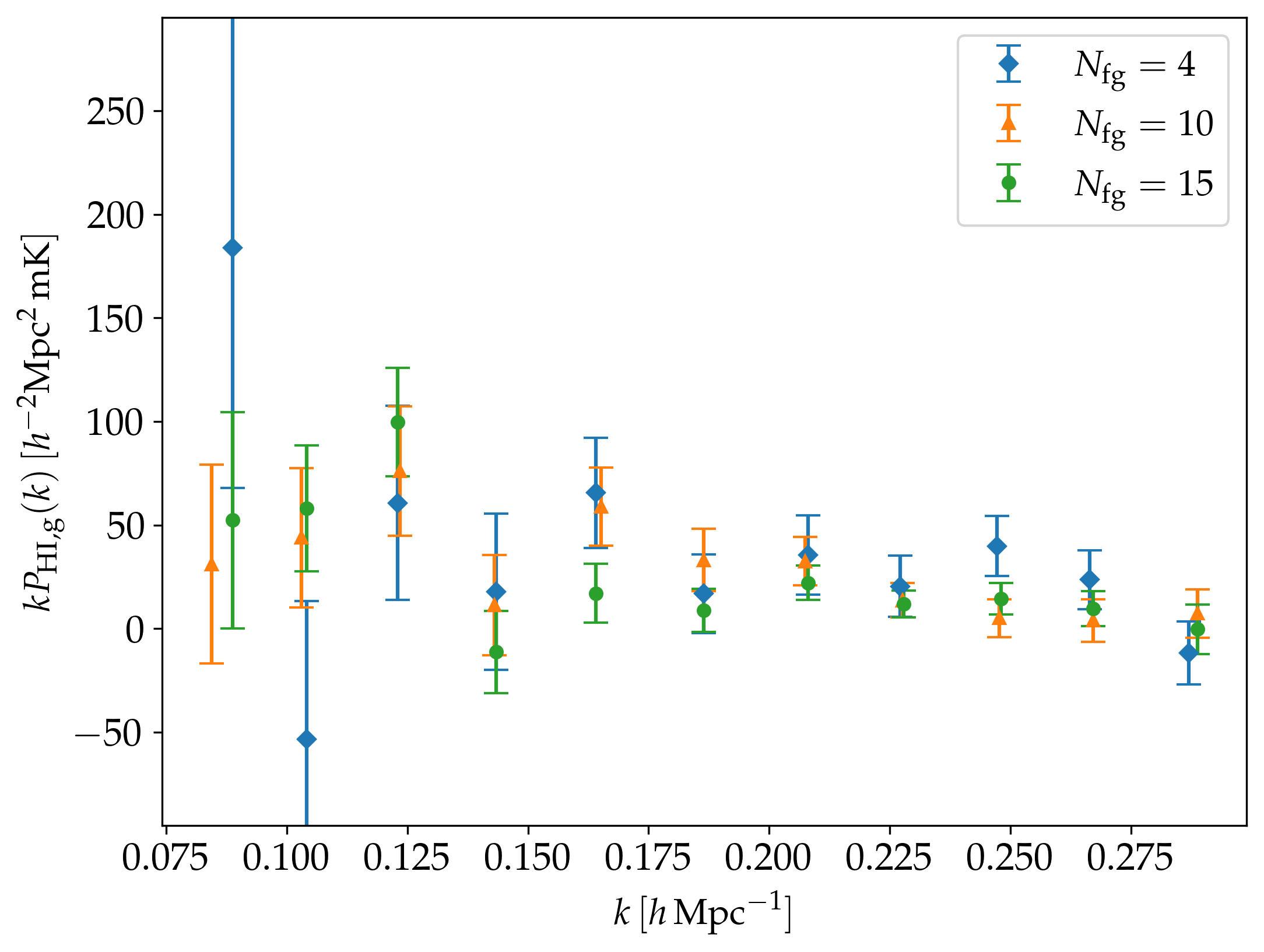}
\caption{\label{fig:nfg_pk}
Cross-correlation power spectrum monopole measured after removing $N_{\mathrm{fg}}=4$, $10$ and $15$ PCA modes from the intensity maps. The largest differences occur in the two lowest-$k$ bins for $N_{\mathrm{fg}}=4$, where the residual foreground power strongly increases the uncertainties. On smaller scales the three measurements are consistent within the uncertainties, with $N_{\mathrm{fg}}=15$ lying below the reference case for $0.14\lesssim k\lesssim0.21\;h\,\mathrm{Mpc^{-1}}$ and $N_{\mathrm{fg}}=4$ lying above it at $k\simeq0.25$--$0.27\;h\,\mathrm{Mpc^{-1}}$.}
\end{figure}

\begin{figure}
\centering
\includegraphics[width=0.5\textwidth]{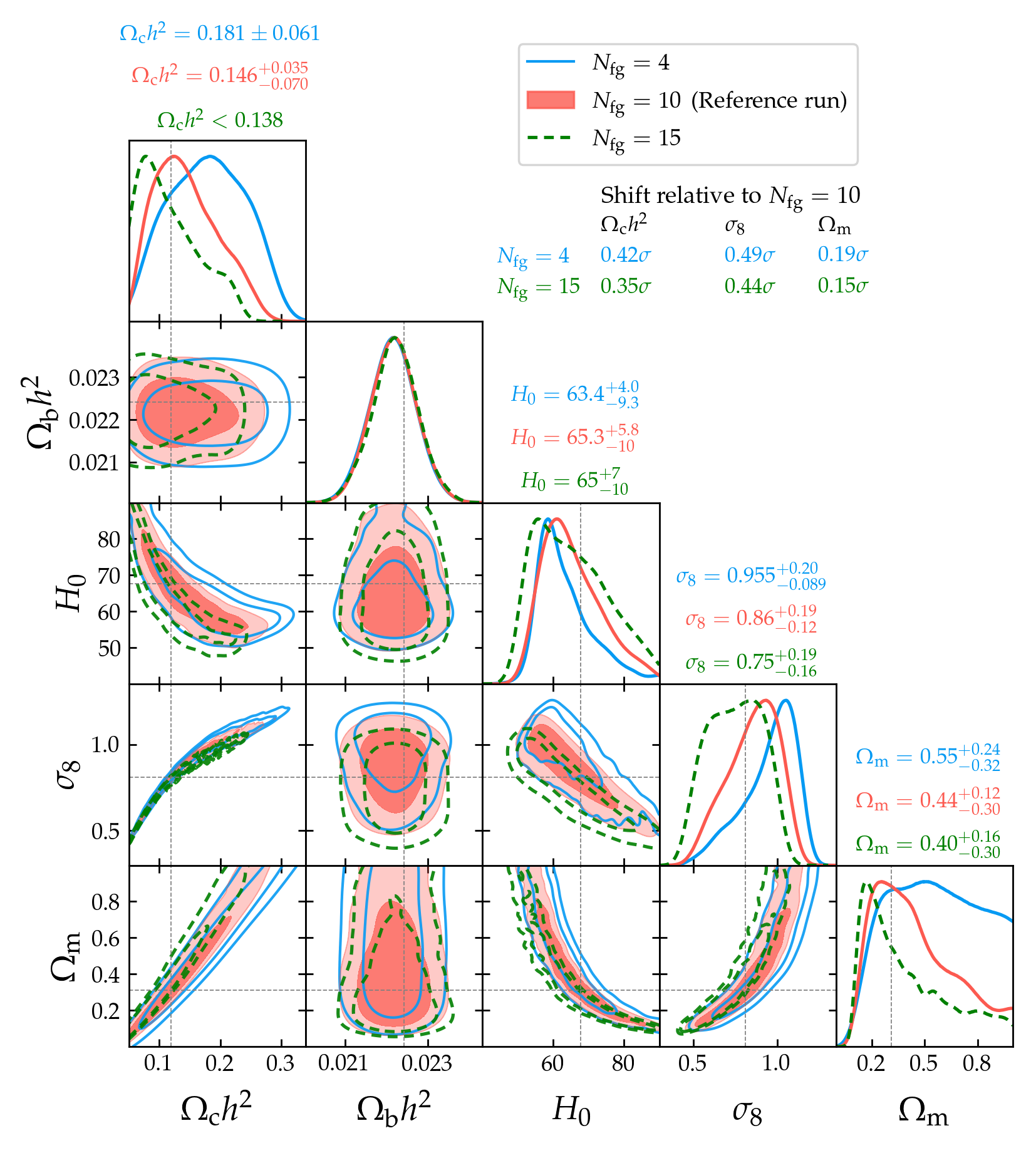}
\caption{\label{fig:nfg_triangle}
Marginalised posteriors obtained by repeating the reference inference on the data vectors cleaned with $N_{\mathrm{fg}}=4$ (blue), $10$ (red, reference run) and $15$ (green, dashed). For $N_{\mathrm{fg}}=15$ the posterior on $\Omega_\mathrm{c}h^2$ reaches the lower edge of the prior, and only an upper limit is quoted. Dashed grey lines indicate the \textit{Planck} 2018 best-fit values.}
\end{figure}

The foreground cleaning removes the $N_{\mathrm{fg}}$ largest principal components of the frequency--frequency covariance of the intensity maps, under the assumption that the foreground emission is spectrally smooth and therefore concentrated in a small number of dominant modes. The choice of $N_{\mathrm{fg}}$ is a compromise, since removing too few components leaves residual foreground power, which contaminates the largest scales and increases the variance of the measurement, while removing too many suppresses a large fraction of the \hi\ signal, in particular the modes with low $k_\parallel$. In the reference analysis we adopt $N_{\mathrm{fg}}=10$, and correct for the resulting signal loss with the transfer function $\mathcal{T}(k)$ of \autoref{eq:transfer_function}. To assess the sensitivity of our results to this choice, we repeat the full analysis on data vectors obtained with $N_{\mathrm{fg}}=4$ and $N_{\mathrm{fg}}=15$, i.e. removing substantially fewer and substantially more components than in the reference case. The transfer function is recomputed for each value of $N_{\mathrm{fg}}$, and all the modelling choices are those of the reference configuration.

The corresponding monopoles are compared in \autoref{fig:nfg_pk}. The largest differences occur in the two lowest-$k$ bins, $k\lesssim0.11\;h\,\mathrm{Mpc^{-1}}$, which are the most affected by the cleaning. For $N_{\mathrm{fg}}=4$ these bins move well away from the other two measurements, but their uncertainties are two to three times larger than in the reference case. This is the expected effect of the residual foreground power left by the milder cleaning. Since the foregrounds are uncorrelated with the galaxy distribution, they do not bias the cross-power spectrum on average, but they enter its variance through the \hi\ auto-power spectrum in \autoref{eq:gaussian_errors}. The same mechanism explains the opposite behaviour at higher $k$, where the uncertainties for $N_{\mathrm{fg}}=15$ are the smallest of the three in most bins. On these smaller scales the three data vectors are consistent within the uncertainties, although individual bins differ by amounts comparable to their error bars. The comparison also illustrates a limitation of the transfer function correction for aggressive cleaning. The correction is a positive multiplicative rescaling of the measured power, and thus it compensates for a suppression of the signal but cannot change its sign. When the cleaned power in a bin is negative, because the signal loss combines with a downward noise fluctuation, the correction amplifies the negative value instead of recovering the underlying signal. 

The posteriors obtained from the three data vectors are compared in \autoref{fig:nfg_triangle}. For $N_{\mathrm{fg}}=4$ we find $\Omega_\mathrm{c}h^2=0.181\pm0.061$, $H_0=63.4^{+4.0}_{-9.3}\;\mathrm{km\,s^{-1}\,Mpc^{-1}}$, $\Omega_\mathrm{m}=0.55^{+0.24}_{-0.32}$ and $\sigma_8=0.96^{+0.20}_{-0.09}$. For $N_{\mathrm{fg}}=15$ we find $\Omega_\mathrm{c}h^2<0.138$, $H_0=65^{+7}_{-10}\;\mathrm{km\,s^{-1}\,Mpc^{-1}}$, $\Omega_\mathrm{m}=0.40^{+0.16}_{-0.30}$ and $\sigma_8=0.75^{+0.19}_{-0.16}$. These are to be compared with $0.146^{+0.035}_{-0.070}$, $65.3^{+5.8}_{-10}\;\mathrm{km\,s^{-1}\,Mpc^{-1}}$, $0.44^{+0.12}_{-0.30}$ and $0.86^{+0.19}_{-0.12}$ in the reference run. The shifts are ordered with $N_{\mathrm{fg}}$, therefore removing fewer modes moves $\Omega_\mathrm{c}h^2$, $\Omega_\mathrm{m}$ and $\sigma_8$ to higher values, and removing more moves them to lower values. For $N_{\mathrm{fg}}=15$ the posterior on $\Omega_\mathrm{c}h^2$ peaks close to the lower edge of the prior, so that only an upper limit can be quoted. For $N_{\mathrm{fg}}=4$ the posterior on $\Omega_\mathrm{m}$ is broad and extends to the upper edge of the allowed range. $H_0$ is the least affected parameter, with posterior means that remain close to each other. 

Since the posteriors are markedly non-Gaussian, we quantify the shifts with the non-Gaussian parameter-shift estimator of \citet{Raveri:2021wfz}, as implemented in \textsc{tensiometer}\footnote{\texttt{https://github.com/mraveri/tensiometer}}. The estimator is applied to the one-dimensional distribution of the difference between each parameter and its value in the reference run. For $N_{\mathrm{fg}}=4$ we find shifts of $0.42\sigma$, $0.49\sigma$ and $0.19\sigma$ in $\Omega_\mathrm{c}h^2$, $\sigma_8$ and $\Omega_\mathrm{m}$, respectively. For $N_{\mathrm{fg}}=15$ the corresponding values are $0.35\sigma$, $0.44\sigma$ and $0.15\sigma$. We do not report shifts for $H_0$ as it is only mildly affected by the change of $N_{\mathrm{fg}}$. It is worth noting that this estimator treats the two chains as independent. The three data vectors, however, are derived from the same maps and are therefore correlated, so the values above may underestimate the significance of the shifts. 

Moreover, the direction of the shifts is consistent with the behaviour of the data vectors in \autoref{fig:nfg_pk}. With $N_{\mathrm{fg}}=4$ the measured power is higher at $k\simeq0.25$--$0.27\;h\,\mathrm{Mpc^{-1}}$, where the uncertainties are small, while with $N_{\mathrm{fg}}=15$ it is lower between $k\simeq0.14$ and $0.21\;h\,\mathrm{Mpc^{-1}}$. Since $A_\mathrm{s}$, the biases and $\Omega_{\mathrm{\hi}}$ are held fixed, a change in the amplitude of the measured power is absorbed mainly by $\sigma_8$ and, through the tight relation between the two parameters, by $\Omega_\mathrm{c}h^2$. Indeed, the three posteriors lie along the same curve in the $\Omega_\mathrm{c}h^2$--$\sigma_8$ plane and are displaced along it. With the present data we cannot establish whether this trend is a statistical fluctuation, driven by the different noise and foreground residuals in the three data vectors, or a residual systematic in the correction for the signal loss, such as an incomplete compensation by the transfer function at large $N_{\mathrm{fg}}$. Since the shifts are below $1\sigma$ for all the parameters, we conclude that, at the precision of the present measurement, the inferred cosmological parameters are robust to the choice of $N_{\mathrm{fg}}$ over the range tested. The reference choice, $N_{\mathrm{fg}}=10$, lies between the two regimes discussed above: it avoids both the large low-$k$ variance of $N_{\mathrm{fg}}=4$ and the unreliable bins of $N_{\mathrm{fg}}=15$. The dependence of the results on $N_{\mathrm{fg}}$ will, however, need to be monitored in future analyses with lower statistical uncertainties.

\subsection{Dependence on the scale cuts}
\label{subsec:scale_cuts}

\begin{figure}
\centering
\includegraphics[width=0.5\textwidth]{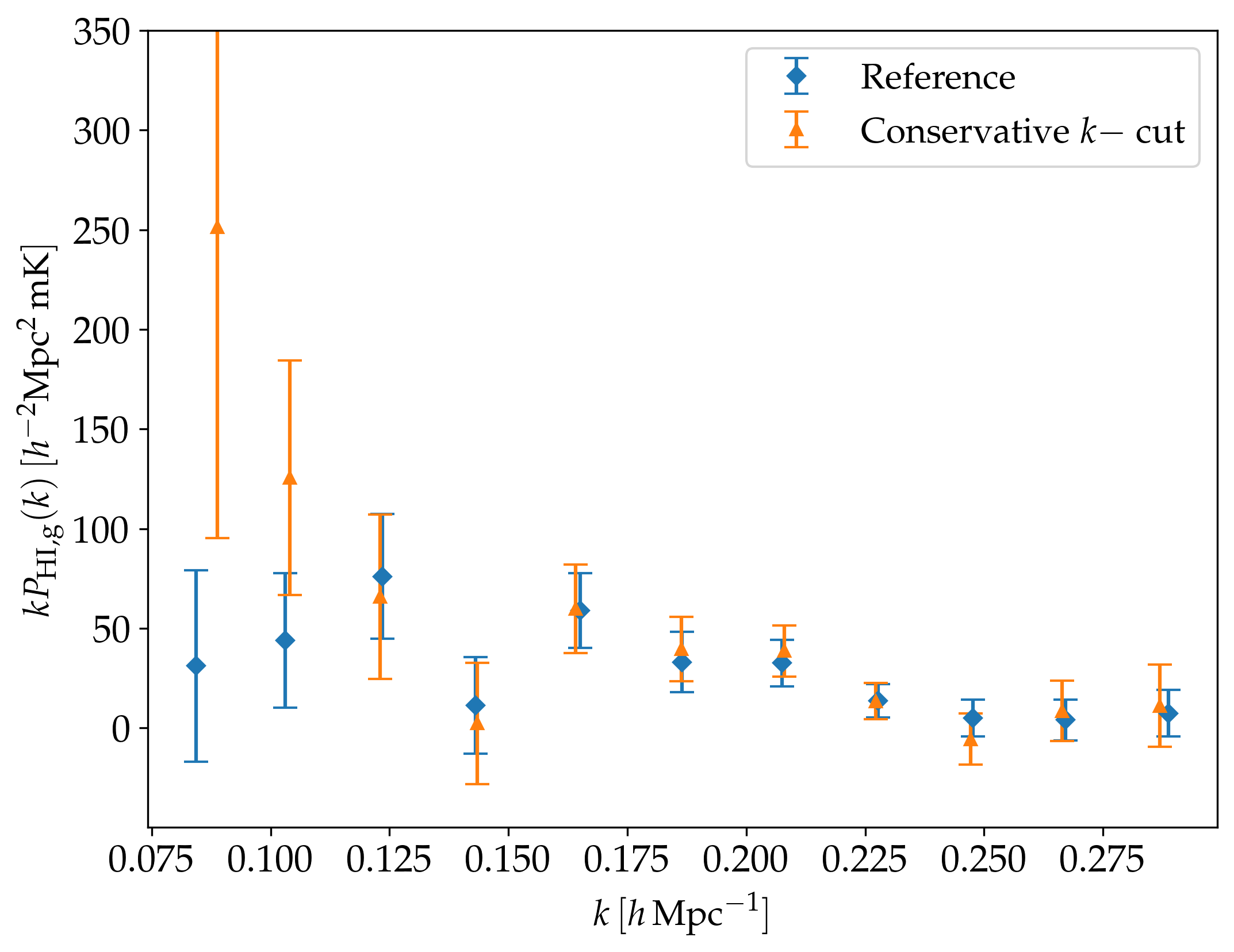}
\caption{\label{fig:kcut_pk}
Cross-correlation power spectrum monopole for the reference selection of modes, $0.02<k_{\parallel,\perp}<0.25\;h\,\mathrm{Mpc^{-1}}$ (blue), and for the conservative selection, $0.05<k_{\parallel,\perp}<0.22\;h\,\mathrm{Mpc^{-1}}$ (orange). Since the cuts are applied separately to $k_\parallel$ and $k_\perp$, both data vectors extend to $k\simeq0.3\;h\,\mathrm{Mpc^{-1}}$. The two lowest-$k$ bins of the conservative data vector contain few modes and have correspondingly large uncertainties.}
\end{figure}

\begin{figure}
\centering
\includegraphics[width=0.5\textwidth]{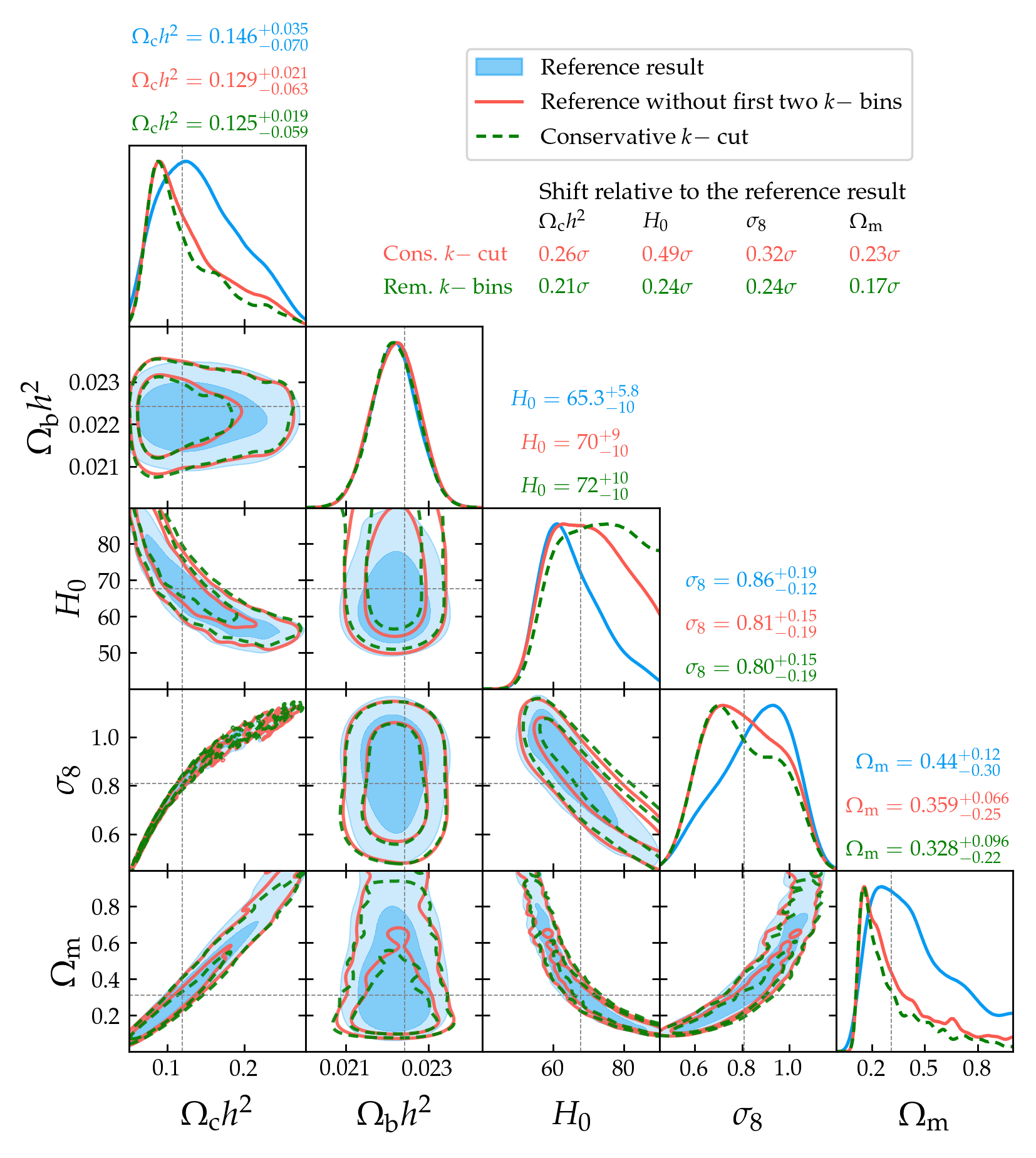}
\caption{\label{fig:kcut_nofirst2kbins_triangle}
Marginalised posteriors for the reference run (blue), for the reference run with the two lowest-$k$ bins removed from the likelihood (red), and for the run using the conservative selection of modes, $0.05 < k_{\parallel,\perp} < 0.22\;h\,\mathrm{Mpc^{-1}}$ (green, dashed). The last two runs give very similar posteriors, showing that the shift of the conservative run with respect to the reference one is driven by the two lowest-$k$ bins. Dashed grey lines indicate the \textit{Planck} 2018 best-fit values.}
\end{figure}

Before being averaged into the one-dimensional bins used in the likelihood, the three-dimensional power spectrum is restricted to a region of the $(k_\parallel, k_\perp)$ plane. The cuts serve different purposes in the two directions. Along the line of sight, the lowest $k_\parallel$ modes are the ones most affected by the foreground cleaning, since a smooth spectral component is by definition a low-$k_\parallel$ feature. The highest $k_\parallel$ modes, instead, are limited only by the channel width, which corresponds to a comoving scale much smaller than any scale of interest here. In the transverse direction the accessible range is set by the angular resolution of the instrument. The beam attenuation of \autoref{eq:beam_attenuation} suppresses the signal exponentially at high $k_\perp$, so that these modes are noise-dominated and their inclusion requires accurate knowledge of the beam.

In the reference analysis we retain the modes with $0.02 < k_{\parallel,\perp} < 0.25\;h\,\mathrm{Mpc^{-1}}$ in both directions. To test this choice, we repeat the analysis with a more conservative selection, $0.05 < k_{\parallel,\perp} < 0.22\;h\,\mathrm{Mpc^{-1}}$. This removes the modes most exposed to foreground residuals at low $k_\parallel$ and those most affected by the beam at high $k_\perp$, at the cost of discarding $\sim 40\%$ of the modes entering the one-dimensional average. The model is evaluated with the same selection of modes, so that the two analyses differ only in the region of the $(k_\parallel, k_\perp)$ plane that is used.

The resulting monopoles are compared in \autoref{fig:kcut_pk}. Since the cuts are applied separately to each direction, the one-dimensional bins still extend to $k\simeq0.3\;h\,\mathrm{Mpc^{-1}}$, but the distribution of modes within each bin changes. The effect is strongest in the lowest-$k$ bins. At $k\lesssim0.1\;h\,\mathrm{Mpc^{-1}}$ most of the modes of the reference selection have either $k_\parallel$ or $k_\perp$ below $0.05\;h\,\mathrm{Mpc^{-1}}$ and are removed. 
These bins therefore contain few modes, and as a result their uncertainties increase by a factor of about two and three. The effective wavenumbers of these bins also shift slightly. In these two bins, the conservative measurement is higher than the reference one, and the differences amount to about $1.5\sigma$ of the conservative uncertainties. At $k\gtrsim0.12\;h\,\mathrm{Mpc^{-1}}$, the two data vectors agree within the uncertainties in every bin, with the conservative uncertainties somewhat larger at the highest $k$ because of the reduced number of modes.

The posteriors are compared in \autoref{fig:kcut_nofirst2kbins_triangle}. With the conservative selection we find $\Omega_\mathrm{c}h^2=0.125^{+0.019}_{-0.059}$, $H_0=72\pm10\;\mathrm{km\,s^{-1}\,Mpc^{-1}}$, $\Omega_\mathrm{m}=0.328^{+0.096}_{-0.22}$ and $\sigma_8=0.80^{+0.15}_{-0.19}$, compared with $0.146^{+0.035}_{-0.070}$, $65.3^{+5.8}_{-10}\;\mathrm{km\,s^{-1}\,Mpc^{-1}}$, $0.44^{+0.12}_{-0.30}$ and $0.86^{+0.19}_{-0.12}$ in the reference run. The posteriors move along the $\Omega_\mathrm{c}h^2$--$H_0$ degeneracy, towards higher $H_0$ and lower $\Omega_\mathrm{c}h^2$, $\Omega_\mathrm{m}$ and $\sigma_8$.

Since the two data vectors differ appreciably only in the two lowest-$k$ bins, we test whether these bins are responsible for the shift. To do so, we repeat the reference analysis after removing them from the likelihood, keeping the reference selection of modes in all the other bins. This run is shown in red in \autoref{fig:kcut_nofirst2kbins_triangle}. We find $\Omega_\mathrm{c}h^2=0.129^{+0.021}_{-0.063}$, $H_0=70^{+9}_{-10}\;\mathrm{km\,s^{-1}\,Mpc^{-1}}$, $\Omega_\mathrm{m}=0.359^{+0.066}_{-0.25}$ and $\sigma_8=0.81^{+0.15}_{-0.19}$. These results are close to those of the conservative run in all the parameters and in all the two-dimensional planes, the largest difference being a slightly more extended tail towards high $H_0$ in the conservative case. This result has two implications. First, the shift of the conservative run with respect to the reference one is due almost entirely to the two lowest-$k$ bins. In the conservative data vector these bins contain few modes and have large uncertainties, so they carry little weight in the likelihood, and the analysis is effectively equivalent to discarding them. Second, removing the modes that the conservative selection discards at $k\lesssim0.12\;h\,\mathrm{Mpc^{-1}}$, including those most affected by the beam, has little effect on the inferred parameters.

The direction of the shift can be understood as follows. With $\Omega_{\mathrm{\hi}}$, the biases and $A_\mathrm{s}$ fixed, the $\Omega_\mathrm{c}h^2$--$H_0$ degeneracy follows from the amplitude of the signal: an increase of $\Tb$ with $H_0$ must be compensated by a lower $\sigma_8$, and hence a lower $\Omega_\mathrm{c}h^2$. Along this direction the shape parameter $\Omega_\mathrm{m}h$ changes monotonically, and the position of the posterior is therefore set by the shape of the measured spectrum. In the reference data vector the two lowest-$k$ bins lie below the trend of the bins at $k\simeq0.12$--$0.17\;h\,\mathrm{Mpc^{-1}}$. This relatively low power on the largest scales disfavours low values of $\Omega_\mathrm{m}h$, which would move the turnover of $P_\mathrm{m}$ to larger scales and steepen its decline over the fitted range. When these bins are removed, or their weight is reduced as in the conservative selection, this constraint is lost. The posterior then extends towards lower $\Omega_\mathrm{m}h$, i.e. towards lower $\Omega_\mathrm{c}h^2$ and higher $H_0$, and $\sigma_8$ decreases slightly, as required to keep the amplitude of the signal fixed while $\Tb$ increases. The shift of the conservative run is somewhat larger, in particular in $H_0$, whose mean moves by $6.7\;\mathrm{km\,s^{-1}\,Mpc^{-1}}$ compared with $4.7\;\mathrm{km\,s^{-1}\,Mpc^{-1}}$ when the two lowest-$k$ bins are removed. This is consistent with the higher values of the two lowest-$k$ bins in the conservative data vector, which pull in the same direction, although with little weight.

The change in the widths of the intervals requires some care. The $68\%$ interval on $H_0$ broadens by $\sim 20\%$ when the two lowest-$k$ bins are removed and by $\sim 27\%$ with the conservative selection, and that on $\sigma_8$ by $8\%$ in both cases, as expected from the reduced information. The intervals on $\Omega_\mathrm{c}h^2$ and $\Omega_\mathrm{m}$, instead, are narrower than in the reference run by $20$--$26\%$, and the corresponding posteriors are more sharply peaked. This does not indicate additional information. Without the constraint from the lowest-$k$ bins, the posterior extends further along the degeneracy direction and reaches the upper edge of the $H_0$ accepted range, $H_0=90\;\mathrm{km\,s^{-1}\,Mpc^{-1}}$, where the marginalised density is still about half of its peak value when the two lowest-$k$ bins are removed and about $85\%$ with the conservative selection. The prior then truncates the degeneracy, which produces a sharp lower edge in the marginalised posteriors of $\Omega_\mathrm{c}h^2$ and $\Omega_\mathrm{m}$, and hence the narrower intervals. For the same reason, the broadening of the $H_0$ interval quoted above is a lower limit. The fact that the peaks of the $\Omega_\mathrm{c}h^2$ and $\Omega_\mathrm{m}$ posteriors occur at the same values in the two runs, $\Omega_\mathrm{c}h^2\simeq0.09$ and $\Omega_\mathrm{m}\simeq0.16$, supports this interpretation.

As in \secref{subsec:nfg}, we quantify the shifts with respect to the reference run with the non-Gaussian parameter-shift estimator of \citet{Raveri:2021wfz}. For the conservative selection we find shifts of $0.26\sigma$ in $\Omega_\mathrm{c}h^2$, $0.49\sigma$ in $H_0$, $0.32\sigma$ in $\sigma_8$ and $0.23\sigma$ in $\Omega_\mathrm{m}$. For the run without the two lowest-$k$ bins we find $0.21\sigma$, $0.24\sigma$, $0.24\sigma$ and $0.17\sigma$, respectively. As discussed in \secref{subsec:nfg}, the estimator treats the posteriors as independent, whereas both data vectors are built from a subset of the information entering the reference one, so these values are likely to underestimate the significance of the shifts. All the shifts are nevertheless well below $1\sigma$, and we regard them as consistent with the removal of part of the data.

Taken together, the tests presented in this section indicate that the constraints of \secref{subsec:main_result} are not driven by a particular set of data analysis choices. Neither the number of foreground modes removed nor the region of the $(k_\parallel, k_\perp)$ plane included in the average changes the inferred parameters significantly compared with the statistical uncertainty. The shifts that we do find can be traced back to specific features of the data vectors. When varying $N_{\mathrm{fg}}$, they follow the amplitude of the power on intermediate and small scales. When varying the range of scales, they are due to the two lowest-$k$ bins, which are also the bins most affected by the foreground cleaning. These two bins have a noticeable influence on the position of the reference posterior along the $\Omega_\mathrm{c}h^2$--$H_0$ degeneracy, since they carry most of the information on the shape of the spectrum that breaks it. This makes a reliable measurement of the largest scales, and hence an accurate treatment of the foregrounds at low $k_\parallel$, particularly important for future analyses. We regard these results as an important element of the proof of concept, since they show that the least well-determined steps of the data analysis can be varied without changing the conclusions of the inference at the present level of precision, and that the origin of the residual shifts can be identified. The same tests will have to be repeated once the uncertainties are reduced, when shifts of the size found here would become significant.

\section{Conclusions}
\label{sec:conclusions}

We have presented an inference of cosmological parameters from the cross-correlation power spectrum between the MeerKLASS 2019 L-band pilot intensity maps and the overlapping WiggleZ galaxies, a measurement that has so far been used only to constrain the amplitude of the \hi\ signal. The full analysis has been carried out with \textsc{meer21cm} \citepaliasyear{2026arXiv260701864C}. The cross-correlation power spectrum model includes the beam, the pixel and channel window functions, the gridding compensation, the AP distortions and the convolution with the survey window and weights. Our aim has not been to obtain competitive constraints, but to show that the analysis can be run end-to-end on real data, that it returns results consistent with the concordance cosmology, and that it responds in the expected way when the assumptions entering it are varied.

As a first check, we have combined the cross-correlation likelihood with the \textit{Planck} 2018 likelihoods, so that the cosmology is fixed by the CMB and the cross-correlation constrains only the amplitude of the signal. We find $\Omega_{\mathrm{\hi}}=(0.85\pm0.17)\times10^{-3}$, consistent within $1\sigma$ with the value obtained by \citet{Cunnington:2022uzo} for the same field. This shows that the amplitude recovered by our pipeline agrees with previous analyses of the same data, and it motivates the value of $\Omega_{\mathrm{\hi}}$ adopted in the reference analysis. From the cross-correlation data alone, with the nuisance parameters held fixed and a BBN prior on $\Omega_\mathrm{b}h^2$, we obtain $\Omega_\mathrm{c}h^2=0.146^{+0.035}_{-0.070}$, $H_0=65.3^{+5.8}_{-10}\;\mathrm{km\,s^{-1}\,Mpc^{-1}}$, $\Omega_\mathrm{m}=0.44^{+0.12}_{-0.30}$ and $\sigma_8=0.86^{+0.19}_{-0.12}$, all consistent with the \textit{Planck} 2018 values within $1\sigma$. These constraints are obtained under the most favourable assumptions about the astrophysical nuisance parameters, and they should be read as a demonstration of the pipeline rather than as a measurement of cosmological parameters.

We have then examined the assumptions entering the model. Sampling $\Omega_{\mathrm{\hi}}$ with a Gaussian prior of $\sim31\%$ width, informed by previous measurements, degrades the constraints only slightly, and parametrising the amplitude through $\Tb\sigma_8$ instead of $\Omega_{\mathrm{\hi}}$ gives consistent results. Including the FoG damping with a wide uniform prior on $\sigma_\mathrm{v}$ leaves the velocity dispersion unconstrained and degrades the constraints substantially, in particular on $\Omega_\mathrm{c}h^2$ and $\Omega_\mathrm{m}$, while a Gaussian prior motivated by hydrodynamical simulations recovers most of the constraining power. In both cases the shifts with respect to the reference run remain within $1\sigma$. Switching off the AP effect changes the posterior means by less than $0.1\sigma$ and broadens the intervals only slightly, as expected for a measurement of the monopole alone, in which the BAO are not resolved. Taken together, these tests show that, at the present precision, the constraints are driven by the shape and amplitude of the power spectrum, and that their interpretation depends on the external knowledge of the amplitude of the \hi\ signal and of the biases, which is the main limitation of the analysis.

Subsequently, we have used synthetic data to quantify the bias that would result from neglecting the observational effects in the model. When all the observational effects are included, the input cosmology is recovered within $1\sigma$ for all the parameters we consider. When they are neglected, the posteriors are biased low by up to $2.5\sigma$, and they are at the same time narrower than the correct ones, so that the bias would not be apparent from the uncertainties alone. Most of this bias is due to neglecting the gridding compensation, whose effect is already appreciable within the range of scales used in the fit, while the pixel and channel window functions have a negligible impact for the present data. This is one of the main results of this work: even for a measurement of limited statistical power, an accurate description of the steps of the data analysis in the model is necessary to obtain unbiased cosmological constraints.

Finally, we have tested the stability of the result against the two least well-determined choices made in the construction of the data vector. Varying the number of foreground modes removed from $N_{\mathrm{fg}}=10$ to $4$ and $15$, and restricting the analysis to a more conservative region of the $(k_\parallel,k_\perp)$ plane, change the inferred parameters by less than $0.5\sigma$ according to the non-Gaussian parameter-shift estimator of \citet{Raveri:2021wfz}. We have been able to trace the shifts to specific features of the data vectors: the amplitude of the power on intermediate and small scales when varying $N_{\mathrm{fg}}$, and the two lowest-$k$ bins when varying the scale cuts. Removing these two bins from the reference data vector reproduces the result of the conservative selection. The same tests have also shown the limitations of the present approach. The transfer function correction cannot recover bins in which the cleaned power becomes negative, which sets a practical limit on how aggressively the maps can be cleaned. In addition, the two lowest-$k$ bins, which are the ones most affected by the foreground cleaning, carry most of the information that breaks the $\Omega_\mathrm{c}h^2$--$H_0$ degeneracy. Since the estimator treats the posteriors as independent, while the data vectors are derived from the same maps, the quoted shifts are likely to underestimate their significance, and a proper assessment would require the joint covariance of the different data vectors.

The analysis presented here relies on some simplifications that will need to be relaxed as the data improve. We have used only the monopole of the cross-power spectrum, a linear model with the nuisance parameters held fixed, and a Gaussian covariance, and we have not included the FoG damping in the reference analysis. Larger data sets, such as the deep L-band field \citep{2025MNRAS.537.3632M} and the ongoing UHF-band survey \citep{Cunnington:2026opd}, and eventually the SKA-Mid, will provide both the statistical power and the volume needed to measure the higher multipoles, to resolve the BAO and to constrain the nuisance parameters jointly with cosmology. The observational effects that we have found to matter here will become more important as the uncertainties decrease, and so will the treatment of the foregrounds on the largest scales. We expect the pipeline and the set of validation tests presented in this work to provide a useful basis for these analyses.

\section*{Acknowledgements}
GA and MV are supported by the INFN INDARK grant.
ZC and AP are funded by a UKRI Future Leaders Fellowship [grant MR/X005399/1; PI: Alkistis Pourtsidou]. SCu is funded by a UKRI Stephen Hawking Fellowship (grant reference EP/U536751/1).

\section*{Data Availability}

The MeerKLASS 2019 L-band data is available at \url{https://meerklass.org/HIdata.html}.
The \textsc{meer21cm} package is available at \url{https://github.com/meerklass/meer21cm}. The data processing and inference scripts can be shared upon reasonable requests to the corresponding authors.



\bibliographystyle{mnras}
\bibliography{bibliography} 




\appendix

\section{Derivation of the covariance matrix}
\label{apdx:cov}
In this appendix, we briefly discuss the covariance matrix calculation outlined in \autoref{eq:gaussian_errors}. We ignore the correlation between different $k$-bins for simplicity. Given the wide $k$-bins we use and the limited signal-to-noise ratio, using only the diagonal covariance has a negligible effects on the final results.

\begin{figure}
\centering
\includegraphics[width=0.5\textwidth]{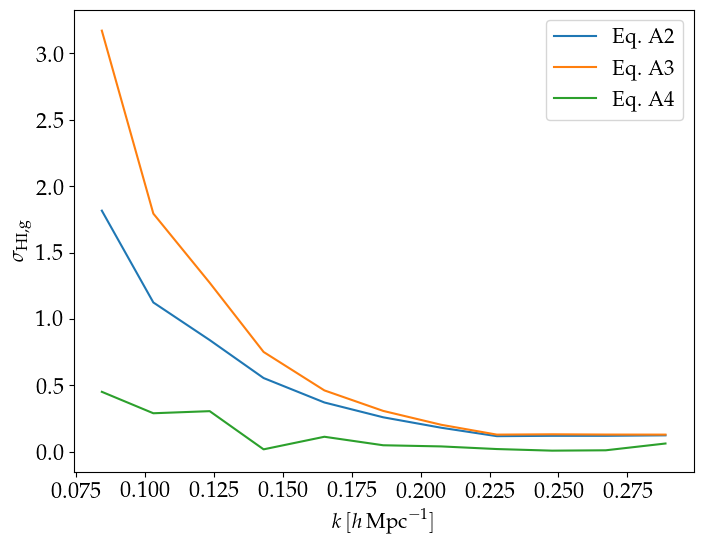}
\caption{\label{fig:covariance_comp}
Comparison of the different methods to estimate the covariance matrix, specifically comparing the methods of \autoref{eq:gaussian_errors}, \autoref{eq:covariance_tf_based} and \autoref{eq:covariance_tf_only}.}
\end{figure}

For the original \hi\ intensity map data vector in Fourier space $\tilde{\bm{T}}_{\rm \hi}$ and the galaxy overdensity field $\tilde{\bm{\delta}}_{\rm g}$, the variance of the cross-power can be written as
\begin{equation}
\begin{split}
    \langle & \hat{P}_{\rm \hi, g}(\bm{k})\hat{P}_{\rm \hi, g}(\bm{k})\rangle - \langle \hat{P}_{\rm \hi, g}(\bm{k})\rangle\langle \hat{P}_{\rm \hi, g}(\bm{k})\rangle
    \\ &= \Big[\big\langle \tilde{\bm{\delta}}_{\rm g}^{*} \tilde{\bm{T}} \tilde{\bm{\delta}}_{\rm g} \tilde{\bm{T}}^{*} \big\rangle -\big\langle\tilde{\bm{\delta}}_{\rm g} \tilde{\bm{T}}^{*} \big\rangle \big\langle\tilde{\bm{\delta}}_{\rm g}^{*} \tilde{\bm{T}} \big\rangle \Big] \bigg/2
    \\&=\Big[ \big\langle\tilde{\bm{\delta}}_{\rm g}^{*} \tilde{\bm{\delta}}_{\rm g} \big\rangle\big\langle\tilde{\bm{T}}^{*} \tilde{\bm{T}} \big\rangle + \big\langle\tilde{\bm{\delta}}_{\rm g}^{*} \tilde{\bm{T}}^{*} \big\rangle\big\langle\tilde{\bm{\delta}}_{\rm g} \tilde{\bm{T}} \big\rangle  \Big] \bigg/2
    \\& = \frac{1}{2}\Big[ \big(P_{\rm \hi,g}^{\rm true}\big)^2 + P_{\rm \hi}^{\rm true}\,P_{\rm g} \Big],
\end{split}
\end{equation}
where we denote the true underlying \hi\ auto-power (including the noise power) as $P_{\rm \hi}^{\rm true}$ and the underlying cross-power as $P_{\rm \hi,g}^{\rm true}$. Note that both the auto- and the cross-power follow $P^{\rm true}=\mathcal{T}^{-1}\langle \hat{P}\rangle$, as shown in \cite{Cunnington:2023ftf} and \cite{2025MNRAS.542L...1C}.
Combined with the assumption of Gaussian variance scaling as $1/{N_{\rm modes}}$ in 1D $k$-bins, the covariance becomes
\begin{equation}
    \sigma^2_\mathrm{H\textsc{i},g}(k) = \frac{1}{2N_{\mathrm{modes}}(k)}\Bigl[P^2_\mathrm{H\textsc{i},g}(k)/\mathcal{T}^2(k) + P_{\mathrm{\hi}}(k)\,P_{\mathrm{g}}(k)/\mathcal{T}(k)\Bigr],
\end{equation}
which is \autoref{eq:gaussian_errors}.
Another way of estimating the covariance is to use the transfer function scatter across different mock realisations. This approach has been used in previous analyses \citep{2018MNRAS.476.3382A, 2025MNRAS.537.3632M} and has been validated in simulations \citep{Cunnington:2023ftf}. The main assumption is that the scatter in the true power comes from the transfer function distribution, $\mathcal{T}(\bm{k}) + \delta\mathcal{T}_i(\bm{k})$, contributed by each $i$th mock realisation. Then, assuming the errors add in quadrature, we obtain
\begin{equation}
    \frac{\delta P_i^{\rm true}(\bm k)}{P^{\rm true}(\bm k)} = \sqrt{\biggl(\frac{\delta P_i(\bm k)}{P(\bm k)}\biggr)^2+\biggl(\frac{\delta \mathcal{T}_i(\bm k)}{\mathcal{T}(\bm k)}\biggr)^2}.
    \label{eq:covariance_tf_based}
\end{equation}
Moreover, if we assume that all quantifiable error lies in the transfer function distribution, therefore $\delta P_i(\bm k)=0$, then 
\begin{equation}
    \delta P_i^{\rm true}(\bm k) =P^{\rm true}(\bm k) \frac{\delta \mathcal{T}_i(\bm k)}{\mathcal{T}(\bm k)}.
    \label{eq:covariance_tf_only}
\end{equation}
We show in \autoref{fig:covariance_comp} the variance of the cross-correlation data computed with \autoref{eq:gaussian_errors} compared to the computation using \autoref{eq:covariance_tf_based} and \autoref{eq:covariance_tf_only}.

We note that, as seen from \autoref{fig:covariance_comp}, the amplitude of the variance used in this work is consistent with the transfer function scatter based method.
The Gaussian covariance sits below \autoref{eq:covariance_tf_based} and above \autoref{eq:covariance_tf_only}.
This is expected, as \autoref{eq:covariance_tf_only} may miss contributions from thermal noise, and \autoref{eq:covariance_tf_based} may overestimate the variance as it ignores correlations between $\delta P_i$ and $\delta \mathcal{T}_i$.
Furthermore, the exact scatter of the transfer function realisations depends on the input model, further complicating the comparison. The method used in this work ensures that we can consistently estimate the covariance with different settings of $N_{\rm fg}$ and $k$-cuts. We leave a detailed study of covariance estimation for future work. 

\bsp	
\label{lastpage}
\end{CJK*}
\end{document}